\documentstyle[aas2pp4,epsfig]{article}

\newcommand{\etal}{ et al. }

\def\gtsim{\ {\raise-0.5ex\hbox{$\buildrel>\over\sim$}}\ }
\def\ltsim{\ {\raise-0.5ex\hbox{$\buildrel<\over\sim$}}\ }

\slugcomment{\today }

\lefthead{LEE \& KIM}
\righthead{GLOBULAR CLUSTERS IN NGC 4472}

\begin{document}

\title{ THE GLOBULAR CLUSTER SYSTEM IN THE INNER REGION OF THE
GIANT ELLIPTICAL GALAXY NGC 4472}

\author{Myung Gyoon Lee and Eunhyeuk Kim}
\affil{Astronomy Program, SEES, Seoul National University, Seoul 151-742, 
Korea \\
Email:  mglee@astrog.snu.ac.kr, ekim@astro.snu.ac.kr}




\begin{abstract}

We present a study of globular clusters in the inner region of the giant
elliptical galaxy NGC 4472, 
based on the $HST$ WFPC2 archive data. We have found about 1560 globular 
cluster candidates at the galactocentric radius $r<4$ arcmin, most of which
are located within $r=1.1$ arcmin. 
$V-(V-I)$ diagram of these objects shows a dominant vertical structure
which consists obviously of two components: blue globular clusters (BGCs)
and red globular clusters (RGCs). 
The luminosity function of the globular clusters is derived to have a 
peak at $V$(max) $=23.50\pm 0.16$ mag and $\sigma = 1.19\pm 0.09$
from Gaussian fitting. 
From this result the distance to NGC 4472 is 
estimated to be $d=14.7\pm1.3$ Mpc, for the foreground reddening
$E(V-I)=0.03$
and the peak luminosity for the Galactic globular clusters
of $M_V$(max) $= -7.4$ mag. 
The peak luminosity for the RGCs is similar to that for the BGCs, 
which indicates
that the RGCs may be several Gyrs younger than the BGCs.
The luminosity function of the globular clusters shows little systematic 
variation depending on the galactocentric radius at $r<4$ arcmin. 
However, the mean luminosity of the bright BGCs decreases by 0.2 mag with increasing
galactocentric radius over the range of 9 arcmin, while that of the RGCs does not. 
The observed color distribution of these globular clusters is 
distinctively bimodal with peaks
at $(V-I) = 0.98\pm0.01$ and $1.23\pm0.01$, 
which correspond to the metallicities of [Fe/H] $= -1.41$ and $-0.23$ dex, 
respectively. 
The mean observed color of all the globular clusters with $V<23.9$ mag is
derived to be $(V-I) =1.11\pm0.01$ (m.e.), 
corresponding to the metallicity of [Fe/H] $= -0.79 \pm 0.05$ dex.
These colors are exactly the same as those for the globular clusters in M87.
It is found that the relative number of the BGCs to
the RGCs is increasing with the increasing galactocentric
radius.
There are more RGCs than the BGCs  within $r \sim 2$ arcmin, while there are
fewer RGCs than the BGCs in the outer region of NGC 4472.
Radial and azimuthal structures of the globular cluster system in NGC 4472
are also investigated.  
Surface number density profiles of both the BGCs and RGCs get flat in the
central region, and the core radii of the globular cluster systems are measured 
to be $r_c = 1.9$ arcmin for the BGCs, $r_c = 1.2$ arcmin for the RGCs, 
and  $r_c = 1.3$ arcmin for the total sample, which are much larger than
the stellar core of the galaxy.
In general the properties of the globular clusters in the inner region of 
NGC 4472 are consistent with those of the globular clusters in the outer region
of NGC 4472.
  
\end{abstract}




%

\section{INTRODUCTION}

NGC 4472 (M49) is a giant elliptical galaxy located at
4$^\circ$ south of the center of the Virgo cluster. 
NGC 4472 is the brightest member of the Virgo cluster, 
some 0.2 mag  brighter than the cD galaxy M87. 
NGC 4472 is an outstanding example of giant elliptical galaxies 
showing a bimodality in the color distribution of globular clusters
(Gebhardt \& Kissler-Patig 1999, Kundu 1999). 
Geisler, Lee, \& Kim (1996) and Lee, Kim, \& Geisler (1998) 
have found from the deep wide
($16.4 \times 16.4$ arcmin$^{2}$) field
CCD imaging of  NGC 4472 that the color distribution of the globular clusters
in NGC 4472 is remarkably bimodal.
The bimodal color distribution of
the globular clusters in NGC 4472 has shown that there are two kinds of
cluster populations in this galaxy: a metal-poor population with a mean metallicity
of [Fe/H] $= -1.3$ dex and a  metal-rich population with a mean metallicity
of [Fe/H] $= -0.1$ dex. Interestingly it is found that the metal-rich
globular clusters show some properties in common with the galaxy halo stars 
in their spatial distribution and color profiles, while the metal-poor globular
clusters do not show such behavior. Also it is found that the metal-rich
globular clusters are more centrally concentrated than the metal-poor globular clusters. 
This result indicates
that there may exist some connection between the metal-rich globular clusters
and halo stars in NGC 4472 in the formation and evolution processes.
However, these studies based on the ground-based observations 
could not investigate the properties of the
globular clusters close to the center of NGC 4472, because of the severe
crowding problem and strong background halo light in ground-based images of
the inner region of the galaxy.  
Hubble Space Telescope ($HST$) data are ideal for the study of the globular clusters 
close to the center of bright galaxies such as NGC 4472.
They are especially useful for the study of the structure of the globular cluster system 
in the central region of NGC 4472 for which little is known by now.

In this paper we present a study of the globular clusters in the inner region
of NGC 4472 using the $HST$ archive data.
During the preparation of this paper 
Puzia \etal (1999) published a paper based on the analysis of the data
which are the same as three sets among the four sets used in this study. 
However, there are several differences between the two studies, 
details of which will be described later.
Here is given a brief summary of the differences between the two studies. 
First, we use one more data set in addition to the data set
used by Puzia et al., 
with which the size of the bright globular cluster sample is
increased by 29\% and the complete areal coverage of the bright 
globular cluster sample is increased significantly
(by a factor of three in the range of the galactocentric radius).
Secondly, we use different methods of data reduction and analysis from 
the Puzia et al.'s. 
As a result, our photometry goes about 1 mag deeper and is more complete than 
the Puzia et al.'s before incompleteness correction. 
Third, $V$ and $I$ magnitudes in our photometry are on average about 0.2 mag fainter 
than the Puzia et al.'s for the common objects, while the colors are almost the same
between the two studies.
Fourth, our results for the peak luminosity of the globular cluster 
luminosity functions are significantly different from the Puzia et al.'s, leading to an opposite conclusion on the age of the globular clusters.
 Fifth, we present for the first time a study of the structure of the central region 
of the globular cluster systems which was not done by Puzia et al.  
Sixth, we use both the $HST$ data of the inner region of NGC 4472 and the
ground-based data of the outer region of NGC 4472 to investigate any systematic
variation over a wide range, 
while Puzia et al. worked with only the $HST$ data of the inner region.

This paper is composed as follows.
Sections 2 describes the data used in this study, and 3 explains the reduction
of the data.
Section 4 presents the color-magnitude diagram of the measured
objects in the NGC 4472 field.
Sections 5 and 6 show, respectively, 
the luminosity functions and color distribution of globular clusters.
Section 7 investigates the spatial structure of the globular cluster system
in NGC 4472. 
Primary results are discussed in Section 8, and
a summary of this study is given in the final section.

\section{DATA} 

We have used four sets of $HST$ Wide Field and Planetary Camera 2 (WFPC2) 
imaging data on NGC 4472 available in
the $HST$ archive for this study: two fields including the center of NGC 4472,
one field 3 arcmin north, and one field 3 arcmin south from the center of NGC 4472.
We name these $HST$ fields: C1 Field, C2 Field, N Field and S Field.
The images were obtained with the wide-band filters F555W and F814W.
Table 1 lists the observation log of these data and Fig. 1 displays
the positions and orientations of the $HST$ fields 
in the digitized Palomar Sky Survey image of NGC 4472.
Areal coverage of the data used in this study 
extends out to $r \approx 4$ arcmin, and is almost complete inside $r = 1.1$ arcmin.

\section{DATA REDUCTION}

\subsection{Image Processing}

C1 and C2 Fields include the nucleus of NGC 4472 so that the background
halo light is much brighter than the globular clusters in the
images. Therefore we need to subtract the galaxy halo 
light before doing photometry of the point sources in the images.  

We first construct smooth models of halo
light using ellipse fitting and median smoothing, then subtract these models from the original images. 
The subtraction works very well and the resulting images show clearly
the point sources
close to the center of NGC 4472. Cosmic rays in the images were removed and individual images of the same 
filter were combined into a single image using CRREJ task in IRAF/STSDAS.

\subsection{Aperture Photometry}

Crowding problem of the point sources seen in the $HST$ images of NGC 4472 is 
negligible so that simple aperture photometry is good
enough to derive reliable photometry of point sources. We have
obtained the photometry of point sources in the images as follows.

First, we find objects with 4 $\sigma$ threshold from the images 
prepared by combining all F555W and F814W images, using FIND in IRAF/DAOPHOT.
Secondly, we derive aperture magnitudes of these objects 
from the combined images of each filter. For this we use CCDCAP
designed for accurate aperture photometry of point sources in the $HST$ images
(\cite{mig98}). We use the aperture radius of 2 pixels, the sky radius
of 5--8 pixels, and the hardness parameter of 0.8 in CCDCAP.
Thirdly, we derive the aperture corrections for the aperture radius of
0.5 arcsec using the bright point sources (which are mostly globular clusters)
in each image,  the results of which are listed in Table 2.
Finally we derive standard magnitudes and colors of these objects
following the description given in Holtzman \etal (1995a, b) and Kundu \etal (1999).
We adopt the zero points of the magnitudes given in Holtzman \etal (1995a, b).
We have tried to measure the excess light of bright 
globular clusters compared with the stellar point spread function 
over the aperture region larger than the radius of 0.5 arcsec, as Puzia et al. did. However, it is found that it
is not possible to measure reliably the amount of the excess light,
because of low signal-to-noise ratios for the outer region.
 The final photometry reaches $V \sim 28$ mag,
and  the total number of objects in the photometry list is about 5670.

It is not easy to secure reliable photometry of the objects in the $HST$
WFPC2 images as ours, especially due to difficulty in measuring the
sky value and deriving the aperture correction. 
Fortunately we can check the accuracy of the photometry using the objects
 located in the overlapped region among the $HST$ fields. 
C2 field (that was not used by Puzia et al. 1999) is overlapped 
with C1 field and N field so that it is very useful
for checking externally 
the accuracy of the photometry of C1 and N fields.
Table 3 and Fig. 2 show the comparison of the photometry for the
point sources common among the $HST$ fields.
In general the photometry of the objects show good agreement among the
overlapped fields, and the mean differences of the photometry for the total sample
are $\Delta V = 0.001 \pm 0.049$ (N = 52) and $\Delta I = -0.026 \pm 0.051$
(N = 59). 
This result shows that the sky measurement and
aperture correction in our photometry are reliable.

\subsection{Image Classification}

There are included both point sources and extended sources 
in the list of the detected objects, and these are foreground Galactic star, 
globular clusters in NGC 4472, or background galaxies.
Here we mean galaxies by extended sources, and mean stars and globular clusters (even
though they are somewhat resolved) by point sources.
Point sources are indeed mostly globular clusters rather than stars, as will be shown
later.
We have selected point sources from the list of the detected objects
using the morphological classifier $r_{-2}$ moment, an indicator for the degree
of central concentration of light of an object (\cite{kro80}).
$r_{-2}$ moment is defined as 
$r_{-2} = [ \sum_i (I_i - sky) /
\sum_i \{(I_i - sky ) / (r_i^2 + 0.5) \} ]^{1/2}$.

We have calculated the values of $r_{-2}$ for $r_i \le 3$ pixels for the detected objects.
The range of $r_{-2}$ defined by the obvious bright point sources are derived 
to be $1.136 < r_{-2}(V) < 1.405$, $1.176 < r_{-2}(I) < 1.465$ for the PC images, and
$0.966 < r_{-2}(V) < 1.286$, and $1.027 < r_{-2}(I) < 1.310$ for the WF images.
The objects with $r_{-2}$ in these range are
classified as point sources.
It turns out that most of the objects in the photometry list are extended
sources.
About 1560 objects among the 5670 detected objects were classified as point sources, 
and the rest as extended source.
These point sources are considered as a final sample of the point sources
for the following analysis.
Photometry of the  bright point sources with $V<22$ mag
are listed in Table 4.
Most of these objects are located at $r<1.1$ arcmin from the center
of NGC 4472, and some of them are located out to $r\approx 4$ arcmin
(in the N Field and S Field).

\subsection{Comparison with Previous Photometry}

We have compared our photometry with the Puzia et al.(1999)'s for the globular
clusters common between the two studies, as displayed in Fig. 3.
Fig. 3 shows that our $V$ and $I$ magnitudes are on average fainter than
the Puzia et al.'s, but that the colors agree very well between the two.
The mean differences between this study and Puzia et al. are 
$\Delta V = +0.14\pm0.05$ mag for 157 objects with $20<V<23$ mag,
$\Delta I = +0.15\pm0.05$ mag for 176 objects with $19<I<22$ mag,
and $\Delta (V-I) = -0.01$.

It appears that there are two causes for the photometric differences between
the two studies which used the same original $HST$ data.
First, part of the difference (0.1 mag) is due to the fact 
that Puzia et al. made a mistake in standard calibration. 
If the Holtzman zero points that are given for the
0.5 radius arcsec aperture are used, 
the additional aperture correction of 0.1 mag
is not needed to derive the total magnitudes (see Holtzman et al. 1995a,b). 
However, Puzia et al. applied this correction, 
making systematically their magnitudes 0.1 mag too bright. 
Secondly, part of the difference (0.04--0.05 mag) may be due to different methods used 
in the two studies. 
We used CCDCAP for the aperture photometry of the point sources 
in the images where the galaxy light is subtracted, while Puzia et al. used Sextractor (Bertin \& Arnouts 1996) for the original images.
It is not unexpected that 0.04--0.05 mag difference exists between the 
different methods of photometry. 
It is also noted that the scatter in the difference in Fig. 3 is larger
than the scatter in the difference among the objects in the overlapped images
in our photometry shown in Fig. 2, indicating that the mean errors
in our photometry are smaller than the Puzia et al.'s.

Puzia et al. derived also additional aperture corrections for the 0.5 arcsec 
radius aperture to the infinite radius aperture for the globular clusters,
using the modeled light profiles:
$C_V = 0.050 \pm 0.015$ mag, $C_I = 0.080 \pm 0.010$ mag, 
and $C_{(V-I)} = 0.030 \pm 0.018$.
Since the modeled light profiles given by Puzia et al. do not match well 
the bright globular cluster profiles at the outer radius (as shown in their Fig. 1),
it is difficult to estimate how reliable the above corrections are. 
Above comparison of the photometry is based on the Puzia et al.'s photometry 
before these corrections are applied. 
If these corrections are included, the magnitude differences 
between the two studies become larger:
$\Delta V = +0.19\pm0.05$ mag, $\Delta I = +0.23\pm0.05$ mag, and $\Delta (V-I) = +0.01$.

\subsection{Artificial Cluster Experiments}

 We have tested completeness of our photometry using artificial cluster
experiment.  We have chosen as a test field 
a field covered by the PC chip for the C1 Field.
This PC field includes the nucleus of NGC 4472 in the center of the image
so that incompleteness for this field is the most severe among the fields
used in this study.

We have derived empirical point spread function of the clusters using  
several isolated brightest clusters in the images and have created 
artificial clusters with different magnitudes by scaling down the images of the 
selected ones. We have created 1000 artificial clusters in  10 images 
for each filter.
Then we have repeated the same reduction
procedure to estimate how many objects are recovered from the artificial
images.
The completeness factors and mean photometric errors obtained in 
this way are displayed in Fig. 4.
Fig. 4 shows that our photometry is complete for $V<24$ mag, 90\% complete
at $V\approx 24.6$ mag, and 50\% complete at $V\approx 25.2$ mag 
(at $I \approx 23.6$ mag) for the region at $10<r<25$ arcsec.
In the innermost region at $r<10$ arcsec, completeness of our photometry
is much lower than the outer region: the completeness is 100\% for 
$V<23.7$ mag, and 50\% at $V\approx 24.5$ mag. 
The photometry of the other fields must be more complete than this PC field,
because the contribution due to varying galaxy light in the other field 
is much weaker than this PC field. 

From this test it is concluded that our photometry is almost complete ($>95\%$)
for $V<24.2$ mag except for the innermost region at $r<10$ arcsec.
We created a final sample of bright globular clusters with $V<23.9$ mag 
used for the analysis of the color of globular clusters, and this sample is
considered to be complete.
The number of the bright globular clusters with $V<23.9$ mag
in the region at $r<10$ arcsec is only
11 so that it is negligible compared with the total sample of bright globular 
clusters used in this study.


\section{COLOR-MAGNITUDE DIAGRAM}

Figs. 5(a) and 5(b) display the $V$--$(V-I)$ diagrams of the measured point
sources (about 1560 objects) and extended objects (about 4110 objects), respectively.
In general, basic features seen in Fig. 5 are similar to those seen
in the color-magnitude diagrams of NGC 4472 obtained from the ground-based
observations by Geisler \etal (1996), but the $HST$ data go much deeper than
the ground-based data.
Fig. 5(a) shows a dominant vertical feature at the color range of 
$0.8< (V-I) < 1.5$, extending up to $V \approx 19.3$ mag.
The objects seen in this feature are mostly globular clusters in NGC 4472.
Contamination due to foreground stars is considered to be negligible for
this feature. 
We consider the point sources with $0.75 < (V-I) < 1.45$  as globular
cluster candidates from now on.
A small number of point sources bluer or redder than this range are probably
not the members of NGC 4472.

The vertical structure in the color-magnitude diagram
appears  obviously to consist of two components: 
one relatively blue component ($0.75<(V-I)<1.08$)
and 
the other relatively red component ($1.08<(V-I)<1.45$). 
We call them, the blue globular clusters (BGCs)
and red globular clusters (RGCs), respectively.
We create a sample of the bright globular clusters with $V<23.9$ mag for
the detailed analysis of the globular cluster system, considering the photometric
errors and the degree of non-member contamination. 
The boxes in Fig. 5 represent the 
boundary of these bright globular clusters. 
The numbers of the blue and red globular clusters with $V<23.9$ mag
are 263 and 346, respectively. 
The relative abundance of the BGCs and RGCs derived for the inner region
of NGC 4472 in this study is opposite to that based on the outer region
of NGC 4472 where there are more BGCs than the RGCs (\cite{geietal96}).
These bright globular clusters are a final sample used for the analysis
of the colors of the globular clusters later.
Fig. 6 displays the positions of the BGCs, RGCs and faint globular cluster candidates with $0.75<(V-I)<1.45$ and $23.9<V<26$ mag.
The property of these two components will be discussed in detail later.

On the other hand, most of the extended objects seen in Fig. 5(b) are fainter
than $V \approx 24.5$ mag. These are mostly faint background galaxies, 
as are often seen in the deep $HST$ images such as the Hubble Deep Field (\cite{wil96}). The striking difference between the color-magnitude
diagram of the point sources and that of the extended sources shows that the image
classification used in this study works very well for the $HST$ data.
A sharp drop in the number of the extended objects seen at $V<  24.5$ mag
in Fig. 5(b) shows 
that there are few globular clusters candidates among the 
extended objects, while a large color scatter in the faint magnitudes seen
in Fig. 5(a) indicates that there are some contamination due to galaxies
in the faint point sources.

We have cross-identified about 190 bright point sources 
with $18.5 < T_1 < 22.0$ mag
common between this study and Geisler \etal (1996).
Fig. 4 displays the relation between $(V-I)$ and $(C-T_1 )$ colors of these
objects.
Using the colors and magnitudes of these objects, 
we have derived transformation relations between $VI$
system and $CT_1 $ system:

$(V-I) = 0.443 (C-T_1) + 0.396 $ for $(C-T_1 ) < 2.62$ with $\sigma = 0.067$,
 
$(V-I) = 1.560 (C-T_1) -2.533 $ for $(C-T_1 ) > 2.62$ with $\sigma = 0.083$, and

$V = T_1 + 0.110 (C- T_1 ) + 0.024 (C-T_1 )^2 + 0.444$ with $\sigma = 0.097$.

Fig. 7 shows the color transformation relation obtained in this study
in comparison with the relation derived from the photometry of the 
standard stars by Geisler (1996):
$(V-I) = 0.514 (C-T_1 ) + 0.115$ and $V = T_1 + 0.256 (C-T_1 ) + 0.052$.
The $(V-I)$ colors derived from the relation obtained in this study is
systematically 
redder than those derived from Geisler's relation 
(by $\sim 0.1$ mag for the color of $(C-T_1) = 1.5$ 
corresponding to the mean color of the
globular clusters). 
We have used the relations derived here 
for combining  the $HST$ $VI$ data in this study
with the ground-based $CT_1$ data
(\cite{geietal96};\cite{lee98}) for further analysis.

In this study we adopt the foreground reddening $E(B-V)=0.0224$ given by
Schlegel, Finkbeiner, \& Davis (1998), while we adopted a zero foreground
reddening given by Burstein \& Heiles (1982) in our previous studies (\cite{geietal96};
\cite{lee98}). For the extinction to reddening ratio $A_V = 3.1 E(B-V)$,
corresponding values are $A_V=0.07$, $A_I=0.03$, and $E(V-I)=0.03$.

\section{GLOBULAR CLUSTER LUMINOSITY FUNCTION}

\subsection{Distance Estimate for NGC 4472}

We have derived the luminosity functions of the globular clusters (GCLF), 
which are listed in Table 5 and displayed in Fig. 8.
In Fig. 8 the luminosity functions of the total sample look almost Gaussian,
with a peak at $V \approx 23.5$ mag. 

We have fit a Gaussian function to the data for the range
 $20.0<V<24.4$ mag, 
for which our photometry is almost complete. 
We derive the peak luminosities: 
$V$(max) = $23.50\pm0.16$ mag and the width, $\sigma = 1.19\pm0.09$,
and $I$(max) = $22.40\pm0.14$ mag and the width, $\sigma = 1.29\pm0.14$. 
Assuming the peak luminosity of the Galactic globular clusters of
$M_V$(max) $= -7.4$ mag (\cite{har96};\cite{lee98}) and
the foreground extinction of $A_V=0.07$ mag, 
we derive a value
for the distance to NGC 4472, $(m-M)_0 = 30.8 \pm 0.2$ and  
$d = 14.7 \pm 1.3$ Mpc.

We have also compared the luminosity functions in this study with the Puzia 
et al.'s in Fig. 9. For this comparison we used the same fields as used
by Puzia et al., C1, N and S Fields and plotted Puzia et al.'s luminosity
functions before incompleteness correction. Also we adjusted the 
Puzia et al.'s magnitudes by +0.14 mag to match our photometry.
Fig. 9 shows that the luminosity functions in the two studies agree well 
for $V<23.5$ mag, but that the luminosity function in this study is somewhat 
higher than Puzia et al.'s for $V>23.5$ mag, especially for the BGCs. 
$I$-band luminosity functions
also show similar results, with some deviations for $I>22.5$ mag. 
This indicates that our photometry is more complete than
Puzia et al.'s before incompleteness correction.
We could not compare the luminosity functions after incompleteness correction,
because the information given in Puzia et al. is not enough for comparison.

\subsection{Comparison of the Peak Luminosity of the BGCs and RGCs}

We have investigated whether there is any systematic difference in the
peak value of the luminosity functions of the BGCs and RGCs.
The $V$-band luminosity functions of the BGCs and RGCs are also listed in 
Table 5 and displayed in Fig. 8.
Fig. 8 shows that the shapes of the luminosity functions of the two
populations are similar and that the peaks are at similar luminosity. 
From Gaussian fitting to the data,
we derive the values for the peak luminosity: 
$V$(max) $= 23.53\pm 0.16$ mag ($\sigma = 1.16 \pm0.11$) for the BGCs,
$V$(max) $= 23.44\pm 0.22$ mag ($\sigma = 1.16 \pm0.13$) for the RGCs,
$I$(max) $= 22.63\pm 0.26$ mag ($\sigma = 1.24 \pm0.20$) for the BGCs, and
$I$(max) $= 22.30\pm 0.17$ mag ($\sigma = 1.32 \pm0.21$) for the RGCs.
Therefore the peak luminosities of the BGCs and RGCs are similar within the
errors.
This result is consistent with the results based on the data of 
the globular clusters in the outer region of NGC 4472 (\cite{lee98}).

The derived values of the sigma for the two populations are exactly the same 
for the $V$-band, and are slightly different for the $I$-band.
If we constrain the sigma as the mean of the two, 1.28, for the $I$-band, 
we get 
$I$(max) $= 22.69\pm 0.13$ mag  and $22.37\pm0.09$ mag, respectively, 
for the BGCs and RGCs. This shows that the difference between the BGCs and
RGCs is changed little by constraining the sigma.

The color chosen as the boundary between the BGCs and RGCs, combined with
observational errors in color, may affect somewhat the luminosity function that
is determined for each group. However, above results  are expected to be 
changed, if any, little, because there is a clear minimum separating
between the BGCs and RGCs in the color distribution, 
and because the total numbers of the clusters 
in each population are not much different. 
We performed a Kolmogorov-Smirnov test for the samples derived
using the color boundary which is 0.05 mag different from the chosen value.
This value of 0.05 mag is five times larger than the mean error of the chosen value.
The Kolmogorov-Smirnov test shows that the two populations derived
using the color boundary which is 0.05 mag different from the chosen value
are the same as the original samples with a probability of 80\%.
Therefore this effect is found to affect little above results.

On the other hand, 
Puzia et al. suggested that the peak luminosities of the BGCs are
0.4--0.5 mag brighter than those of the RGCs:
$V$(max) $= 23.62\pm 0.09$ mag and  $I$(max) $= 22.48\pm 0.07$ for the BGCs, 
and
$V$(max) $= 24.13\pm 0.07$ mag and $I$(max) $= 22.90\pm 0.11$ for the RGCs.
If we adjust Puzia et al's magnitudes to match our photometry as shown before,
these values will be changed to 
$V$(max) $= 23.76$ mag and  $I$(max) $= 22.63$ for the BGCs, and
$V$(max) $= 24.27$ mag and $I$(max) $= 23.05$ for the RGCs.
Therefore the peak luminosities for the BGCs in this study are similar to
the Puzia et al.'s, but the peak luminosities for the RGCs in this study
as much as 0.8 mag brighter than the Puzia et al.'s.

The causes for this large difference for the RGCs between the two studies 
are not clear.
However, there are a few reasons that our results are considered to be more reliable than the Puzia et al.'s.
First, the peak luminosity differences between the BGCs and RGCs given by
Puzia et al. are significantly different between the C1 Field and (N+S) Fields: 
$\Delta V = -0.42$ mag and $\Delta I = -0.21$ mag for the C1 Field, and
$\Delta V = -0.68$ mag and $\Delta I = -0.64$ mag for the (N+S) Fields.
This indicates that there may be some problems in the data even after the 
incompleteness correction of their sample.
Secondly, our sample of globular clusters is about 30\% 
larger than Puzia et al.'s
(because we included the C2 field which was not used by Puzia et al.).
Thirdly, our photometry is more complete than Puzia et al.'s. 
   
Puzia et al.  used further these differences in magnitudes between the BGCs and RGCs
to investigate any age differences between the two components. 
From the comparison of the peak luminosities of the BGCs and RGCs with
the simple stellar population models (Worthey 1994; Bruzual and Charlot 1996; Maraston 1998; Kurth et al 1999) Puzia et al. concluded that 0.4 -- 0.5 mag 
differences between the two components indicate that
the BGCs and RGCs are coeval within the errors of $\sim 3$ Gyrs.
However, our results indicate that the RGCs are 
 several Gyrs younger than
the BGC, if we use the same models used by Puzia et al. (their Figures
9, 10, 11 and 12). 

Our result on the age difference is similar 
to the results for the globular clusters 
in the inner region of M87 (Kundu et al. 1999). 
Kundu et al.  (1999)
found from the analysis of the WFPC images of the inner region of M87 
that the BGCs are 0.2--0.3 mag brighter than the RGCs at $V$-band,
and that the RGCs are 0.06 mag brighter than the BGCs at $I$-band, and
suggested that the RGCs may be 3--6 Gyrs younger than the BGCs.

\subsection{Spatial Variation of the Luminosity Function}

Globular clusters in the inner region of giant galaxies are ideal targets
to study the dynamical effects on the globular cluster system due to
external processes. 
Primary external processes driving dynamical evolution of globular clusters
are tidal shocks and dynamical friction.
According to the theoretical prediction, 
the most massive clusters will be preferentially depleted by dynamical
friction, while less massive clusters with lower density will be more
efficiently destroyed by tidal shocks. 
 These external dynamical effects are expected to be strongly enhanced 
in the inner region of the galaxies 
(\cite{ost97};\cite{gne97};Murali \& Weinberg 1997a,b,c).
As results of these effects, it is expected that 
the globular cluster luminosity function for the inner region of a galaxy
will get flatter than that for the outer region, and the peak luminosity
for the inner region will get brighter than that for the outer region. 
However, several observational studies found little evidence supporting
these theoretical predictions in the case of the globular clusters 
of M87 (\cite{har98};\cite{kun99}).

NGC 4472 is another good galaxy to study this aspect.
We have investigated whether the globular cluster luminosity function 
of NGC 4472 shows any systematic variation on the galactocentric distance,
using three methods: a) the slope of the bright globular cluster luminosity
function, b) the peak luminosity of the globular cluster luminosity function, and
c) the mean luminosity of the bright globular clusters.
Fig. 10 displays the globular cluster luminosity function for five
radial bins. 
We have fit the data for the bright part in the range of $21.5<V<23.5$ mag
using a logarithmic line to check any systematic variation in the slope, 
the results of which are shown by the dashed line.
There is seen little systematic variation depending on the galactocentric radius
in the logarithmic slope of the bright part of the luminosity function.
Next we display the peak luminosity of the  $V$ and $I$ bands 
globular cluster luminosity function for each radial bin in Fig. 11.
Fig. 11 shows little systematic variation of the peak luminosity depending
on the galactocentric radius within $r=2.5$ arcmin. 
The peak luminosity at the outermost position
of $r = 3.3$ arcmin seems to be 0.3 mag fainter than the mean value shown
by the dashed line in Fig. 11. 
This may be either an intrinsic property
of the globular clusters or may be caused by the fact that 
the degree of field contamination due to background galaxies is increasing
in the outer region (see the asymmetric globular cluster luminosity function
for $2'.67 < r < 4'.00$ in Fig. 10).  
This point needs to be checked with similar $HST$ 
observations for larger areas in NGC 4472.
However, the following result indicates that the first possibility is
low.

We have calculated the mean luminosity of the bright globular clusters
with $20<V<23.5$ mag. The magnitude range is chosen that the resulting
mean luminosity is not affected at all by incompleteness in our photometry.
Fig. 12 displays the mean luminosity of the globular clusters versus 
galactocentric radius based on the $HST$ data, and Table 6 lists
the linear fits to the data.
Fig. 12 shows that there is little, if any, change in the mean luminosity
of the bright globular clusters for the region at $r<4$ arcmin from the center
of NGC 4472 (the determined values of the slope are smaller than the errors).
This result is similar to that of the globular clusters in the inner region
of M87 (\cite{har98};\cite{kun99}).

Then we have combined these $HST$ data with those for the outer
region of NGC 4472 given by Geisler et al. (1996) to investigate
any gradient of the mean luminosity 
over a large range of radius, the results of which
are illustrated in Fig. 13. 
Fig. 13 shows that the mean luminosity
of the BGCs is clearly decreasing  by about 0.2 mag over the range of
$r=9$ arcmin (the slope = $0.024 \pm 0.010$ for $V$ and $I$), 
while the mean luminosity
of the RGCs is almost constant over the same range of radius.
This result of the difference between the BGCs and RGCs is not affected by the 
errors involved with combining the $HST$ data and the ground-based data,
because it is based on the globular clusters with the same magnitude range.   
 
Therefore it is concluded that there is little systematic variation of the
globular cluster luminosity function within the range of $r=4$ arcmin,
and that the BGCs on average become  fainter by 0.2 mag with the increasing galactocentric radius over the range of 9 arcmin, while the RGCs
do not. This result shows that the external dynamical effects are much weaker
than the theoretical predictions for the globular cluster systems in NGC 4472. 
The cause for the 0.2 mag variation of the mean luminosity seen 
only for the bright BGCs is unclear. If it is due to dynamical effects, 
it may indicate that the BGCs are older than the RGCs. 
   
\section{GLOBULAR CLUSTER COLOR DISTRIBUTION}

We have derived a color distribution of the bright globular clusters with
$V<23.9$ mag in NGC 4472, which is listed in Table 7 and displayed in Fig. 14.
Fig. 14 shows that the color distribution of the globular clusters
is clearly bimodal. From double-Gaussian fitting to the data, 
we have derived the peak colors  $(V-I)=0.975\pm 0.005$ with $\sigma =0.076$,
and $(V-I)=1.233\pm 0.006$ with $\sigma =0.094$. These results agree
very well with those given for $V<23.75$ mag by Puzia \etal (1999):
$(V-I)=0.99\pm 0.01$ and $1.24\pm 0.01$, respectively. 

Using the relation between $(V-I)$ and metallicity of Galactic globular
clusters given by Kundu \& Whitmore (1998), [Fe/H] $= -5.89 + 4.72 (V-I)$,
and adopting the reddening of $E(V-I)=0.03$, 
we derive the peak metallicities
[Fe/H] $= -1.41$ and --0.23 dex       
for the BGCs and RGCs, respectively, from the peak colors. 
(If we use the slightly different relation used by Harris \etal (1998),
[Fe/H] $= -6.67 + 5.56 (V-I)$, we derive 
[Fe/H] $= -1.39$ and +0.00 dex        
for the BGCs and RGCs, respectively.)
The mean color of the bright globular clusters with $V<23.9$ mag is
derived to be $(V-I) =1.11\pm0.01$ (m.e.), 
corresponding to the metallicity of [Fe/H] $= -0.79 \pm 0.05$ dex.
These results for the peak metallicity 
are in good agreement with the results for the globular
clusters in the outer region of NGC 4472 (\cite{geietal96}; \cite{lee98}).

However, the relative abundance of the BGCs and RGCs is opposite between
the inner region (this study) and the outer region (\cite{geietal96}). 
In the inner region, the RGCs are more
numerous than the BGCs, 
while it is opposite in the outer region.
Fig. 15 displays a radial variation of the color distribution of the
bright globular clusters. It shows that the relative number of the BGCs
to that of the RGCs is increasing as the galactocentric distance
is increasing. 

In Fig. 16 we display a radial variation of the number of the BGCs to that of the RGCs (= N(BGC)/N(RGC)). We also plot the data of the globular clusters
with $T_1 < 23$ mag in the outer region of NGC 4472
derived from Geisler \etal (1996). Fig. 16 shows that the radial variation 
of the ratio is consistent between the inner region and the outer region.
The ratio N(BGC)/N(RGC) increases continuously out to the limit of the data,
$\sim9$ arcmin.

\section{SPATIAL STRUCTURE OF THE GLOBULAR CLUSTER SYSTEM}

We have investigated the spatial structure of the globular cluster system
in NGC 4472 using the sample of the bright globular clusters
with $V<23.9$ mag. 
Without the $HST$ data it is very difficult to study the spatial
structure of the globular cluster systems in the central region of 
NGC 4472.
Fig. 17 displays the spatial positions of these objects in the central
$2.4 \times 2.4$ arcmin$^2$ region of NGC 4472.
It is seen immediately in Fig. 17
that the RGCs are more centrally concentrated than the BGCs.
Radial and azimuthal structures of the globular cluster system are
studied in the following.

\subsection{Radial Structure}

\subsubsection{Surface Number Density} 

We have derived radial profiles of the surface number density of the globular
 clusters with $V<23.9$ mag, which are listed in Table 8 
and are displayed in Fig. 18. 
Fig. 18 shows 
(a) that the surface density increase with decreasing
radius,
(b) that the surface density profiles of both the RGCs and BGCs 
get flat in the central region, and
(c) that the surface density  profile of the RGCs is steeper than that of the BGCs.

The surface density profiles
within $r \sim 2 $ arcmin are approximately fit by King models(\cite{kin66}):
 core radius $r_c = 1.3$ arcmin (= 6.0 kpc), concentration parameter 
$c = \log ( r_t / r_c ) = 3.0$ (where $r_t$ is a tidal radius), and peak surface density
$\sigma_{GC}(0) = 81.3 $ clusters arcmin$^{-2}$ for the total sample,
$r_c = 1.9$ arcmin (= 8.7 kpc), $c = 2.25$, and $\sigma_{GC}(0) = 26.0 $ clusters arcmin$^{-2}$ for the BGCs, and $r_c = 1.2$ arcmin (= 5.5 kpc), $c = 2.25$, and $\sigma_{GC}(0) = 51.3 $ clusters arcmin$^{-2}$ for the RGCs.
The core radius of the globular cluster system derived here is much larger
than that of the stellar halo, $r_c = 3.6 $ arcsec (= 280 pc) given by Kim, Lee
\& Geisler (2000).


In Fig. 19 we display both the surface density profiles
 for the globular clusters in the inner region derived in this study
and those for the globular clusters with $T_1 < 23.0 $ mag in the outer region
 given by Lee \etal (1998). 
The lower limit for the outer globular cluster sample, $T_1 = 23.0$ mag,
corresponds to $V \approx 23.7$ mag  
for the mean color of the globular clusters. 
We plot the surface density profile
of the inner globular clusters with $V<23.7$ mag in Fig. 19.
Note that the surface density profiles of both data agree approximately
for the overlapped region at $1<r<3$ arcmin.
The surface density profiles of the combined data of the outer region at $r > 55$ arcsec 
are well fit by de Vaucouleurs law:

$\log \sigma_{GC} = -1.318(\pm 0.048) r^{1/4} + 2.905(\pm 0.062)$ for the
total sample,
 
$\log \sigma_{GC} = -0.890(\pm 0.058) r^{1/4} + 2.087(\pm 0.076)$ for the
BGCs, and

$\log \sigma_{GC} = -1.778(\pm 0.063) r^{1/4} + 3.148(\pm 0.080)$ for the
RGCs,

\noindent where $r$ is given in units of arcmin. 

\subsubsection{Color}

We have derived the radial profiles of the mean and median 
color of the globular clusters, 
which are listed in Table 9 and displayed in Fig. 20. 
We also displayed the radial profile
of the color of the halo of NGC 4472 in Fig. 20 to compare the globular
clusters and halo of NGC 4472.
We have transformed the $(C-T_1 )$ color of the halo of NGC 4472
(Kim, Lee \& Geisler 2000) into the $(V-I)$ color
using the relation derived in this study.
Fig. 20 shows 
(a) that the mean color of the total sample of globular clusters
decreases with increasing radius for $r>1$ arcmin, 
(b) that the mean colors of the BGCs and RGCs change little with
increasing radius, and (c) that the profile of the RGCs is very similar
to that of the halo. These results are consistent with those for the outer
region given in Lee \etal (1998).

Fig. 21 shows the radial profiles of the metallicity of the globular
clusters derived from the colors based on this study (filled circles) 
and Geisler \etal (1996) (open circles). 
The mean metallicity is increasing with decreasing galactocentric radius. However,
the metallicity becomes almost constant within $r \sim 1$ arcmin, where
the surface number density of the globular clusters is almost constant
as shown in Fig. 18.
 
Interestingly Harris \etal (1998) also
pointed out a similar trend from the ground-based data for M87: 
the mean metallicity of the globular clusters in M87 is almost constant for $r<1$ arcmin. 
However, Kundu \etal
(1999) mentioned that their observations of the inner region of M87
based on the $HST$ data 
do not provide any compelling supporting evidence for the claim 
by Harris \etal (1998).

The metallicity profile for the combined sample of the inner region
and outer region ($r>50$ arcsec) is fit by 
[Fe/H]$= - 0.390(\pm0.054) \log r + 0.120(\pm0.124)$ with $\sigma = 0.037$, where
$r$ is given in terms of arcmin. 
This result is very similar to that
derived for the outer region, shown by the dashed line with a slope of
--0.41 in Fig. 21 (\cite{geietal96}).

\subsection{Azimuthal Structure}

We have investigated the azimuthal distribution of the globular clusters.
Fig. 22 displays azimuthal variations of the globular cluster number
density for $r<1.1$ arcmin where the areal coverage of the $HST$ data
is complete. It shows that the total sample has obvious peaks at a position
angle of 160 and 340 deg and that the RGCs have a peak at 160 deg,
while the BGCs show an almost uniform distribution. The presence of only
one  peak instead of two for the RGCs is due to the asymmetric
distribution of the globular clusters extended along the south-east
direction, as shown in Fig. 17. 

\section{DISCUSSION}

\subsection{Comparison of the Globular Clusters in the Inner and Outer
Regions of NGC 4472}

In this study we have investigated several aspects of the globular clusters in the inner
region of NGC 4472: color-magnitude diagram, luminosity function, color distribution,
and spatial structure of the globular cluster system.
$HST$ data enable us to study the globular clusters located close to the nucleus
of NGC 4472, which was difficult in the case of ground-based data.

We have compared these results to those of the globular clusters in the outer region of 
NGC 4472 (\cite{geietal96};\cite{lee98}), finding that
the properties of the globular clusters in the inner region 
and the outer region of NGC 4472 are in general similar.
The only exception is that there are more RGCs than the BGCs in the inner region, 
while it is opposite in the outer region.
However, this is also
expected from the relation of N(BGC)/N(RGC) and galactocentric distance derived
from the data of the outer region.

Implications of the results based on
 the globular clusters in the outer region of NGC 4472
for the origin of the globular clusters 
were discussed in detail in Lee \etal (1998), 
which remain still valid with the new
results for the globular clusters in the inner region of NGC 4472.
In summary, the observed properties are consistent with many
of the predictions of both the model of episodic in situ formation plus
tidal stripping of globular clusters given by Forbes, Brodie \& Grillmair (1997), and
the gaseous merger model given by Ashman \& Zepf (1992), but each of the models
also has some problems. 

\subsection{Comparison of the Globular Clusters in the Inner Regions of NGC 4472 and M87}

NGC 4472 and M87 (NGC 4486) are the two most dominant galaxies in the Virgo cluster.
NGC 4472 is the brightest member of the Virgo cluster, but not a cD galaxy, while M87 is
a cD galaxy at the center of the Virgo cluster and is only slightly fainter than
NGC 4472. 
M87 has been a target of much more studies than NGC 4472. 
Recently Kundu \etal (1999) presented a study of 1057 globular clusters
in the inner region ($r<1.5$ arcmin) based on the $HST$ WFPC2 data of a field
centered on the nucleus of M87. The areal coverage of their data is
similar to that of C1 and C2 Fields in our study.
Kundu \etal adopted foreground extinctions of $A_V = 0.067\pm 0.04$
and $A_I = 0.032\pm0.02$ in their study of M87,  which are the same adopted
for NGC 4472 in this study. 
We have compared the properties of the globular clusters 
in the inner region of these two galaxies based on the similar data (this study
and \cite{kun99}). 


The luminosity function of the M87 globular clusters has a peak at
$V_0$(max) $ = 23.67\pm0.07$ mag with $\sigma = 1.39\pm 0.06$, which are
slightly larger than that of NGC 4472, $V_0$(max) $ = 23.43\pm0.08$ mag 
with $\sigma = 1.19\pm 0.09$.
Thus NGC 4472 is considered to be slightly closer than M87.
The luminosity function of the M87 globular clusters does not show any significant
spatial variation over the radius of 1.5 arcmin, similarly to the case of NGC 4472.

The color distribution of the globular clusters in M87 is also bimodal. The
peak colors are $(V-I)_0 = 0.95$ and $1.20$ and the mean color of the total
sample is $(V-I)_0 = 1.09$. 
These values are essentially the same as those
of the globular clusters in NGC 4472, $(V-I)_0 = 0.95$, 1.20 and 1.08.
Thus the characteristic colors of the globular clusters in NGC 4472 and M87 are
almost the same.

The surface number density profiles of the globular clusters are flat in the central
regions of both M87 and NGC 4472. 
The core radii of the globular clusters in the inner region of M87 and the stellar halo are, respectively, 56 arcsec and 6.8 arcsec (Kundu et al. 1999). 
Therefore the core of the globular cluster system
in M87 is somewhat smaller than that of NGC 4472, 78 arcsec, 
while the core of the stellar halo in M87 is twice 
larger than that of NGC 4472, 3.6 arcsec.
The relative number of the BGCs to that of the RGCs in M87 is also increasing with increasing galactocentric radius, as in the case of NGC 4472.
The absence of the spatial variation of the globular cluster luminosity function
and the  core radius of the globular cluster systems larger than the
stellar core of the galaxies are considered to support the suggestion that
the large core of the globular cluster systems is a relic of the cluster formation
process rather than a result due to destruction through tidal shocking and 
dynamical friction (Grillmair, Pritchet, \& van den Bergh 1986, Harris \etal 1998).

However, the surface central number density of the globular clusters 
is about six times larger in M87  (460 clusters arcmin$^{-2}$) than in NGC 4472
(81 clusters arcmin$^{-2}$). This indicates that 
the conditions of formation process of globular clusters might have been 
different between M87 and NGC 4472. 

In summary, 
we conclude that the properties of the globular clusters in the inner 
regions of NGC 4472 and M87 are very similar in general.
The only significant difference is that  the surface central number density 
of the globular clusters (and the specific frequency of the globular
clusters correspondingly) is much higher in M87 than in NGC 4472. 


\section{SUMMARY AND CONCLUSION}

We have presented $VI$ photometry of the globular cluster candidates
in the inner ($r<4$ arcmin) region of NGC 4472, derived from the
$HST$ WFPC2 archive data for the four fields of NGC 4472.
We have found about 1560 globular cluster candidates from the colors
and image classifiers. Primary results and conclusions obtained in this study
 are summarized as follows.

\begin{enumerate}

\item The color-magnitude diagram of the measured point sources shows
a dominant vertical structure at the color range of $0.75<(V-I)<1.45$, 
most of which are globular clusters in NGC 4472.
This structure consists obviously of two populations: the blue globular
clusters and the red globular clusters.

\item The luminosity function of the globular clusters is well fit by
a Gaussian function with a peak 
at $V=23.50\pm0.16$ mag and $\sigma = 1.19\pm0.09$. 
From this the distance to NGC 4472 is derived to be
$d=14.7\pm1.3$ Mpc. The globular cluster luminosity function shows
little systematic variation depending on the galactocentric radius.
There is little difference in the peak luminosity between the
BGCs and RGCs. This indicates that the RGCs may be several Gyrs younger than the BGCs,
contrary to the conclusion given by Puzia \etal (1999). 

\item The color distribution of the 609 bright globular clusters with $V<23.9$ mag is
distinctively bimodal with peaks at $(V-I)=0.975$ and $1.233$.
The metallicities corresponding to these peaks are [Fe/H] $= -1.41$ and
$-0.23$ dex.
The mean color of the bright globular clusters with $V<23.9$ mag is
derived to be $(V-I) =1.11\pm0.01$(m.e.), 
corresponding to the metallicity of [Fe/H] $= -0.79 \pm 0.05$ dex.

\item The ratio of the number of the BGCs to that of the RGCs is increasing
with increasing galactocentric radius from the center out to about 9 arcmin.
Thus the RGCs are more centrally concentrated than the BGCs.

\item The surface number density profile of the globular clusters in the
inner region is approximately fit by a King model with core radius $r_c = 1.3$
arcmin, while that of the globular clusters in the outer region is better
fit by de Vaucouleurs law.
The core radius of the globular cluster system obtained in this study is
much larger than that of the stellar halo.

\item The mean colors of the BGCs and RGCs change
little with increasing radius, 
and the mean color of the RGCs is very similar to that
of the stellar halo of NGC 4472.

\item The radial gradient of the metallicity of the entire sample of the
globular clusters with $V<23.9$ mag is approximately fit by 
[Fe/H] $ = - 0.390 \log r$  $+ 0.120$ with $\sigma = 0.037$,
where $r$ is given in the unit of arcmin. 

\item In general the property of the globular clusters in the inner region of 
NGC 4472 is consistent with that of the globular clusters in the outer region
of NGC 4472, except that there are more RGCs than the BGCs in the inner region.

\end{enumerate}

\acknowledgments

The authors are grateful to Sang Chul Kim for careful reading of the
manuscript of this paper and 
to Thomas Puzia for providing the photometry data of 
the globular clusters of NGC 4472 used in Puzia et al. (1999) and
useful information.
Anonymous referee is thanked for useful comments.
This research is supported 
in part by the KOSEF/KISTEP International Collaboration Research Program
(1-99-009).  
 

%

%
%

\clearpage

\begin{figure}
\epsfig{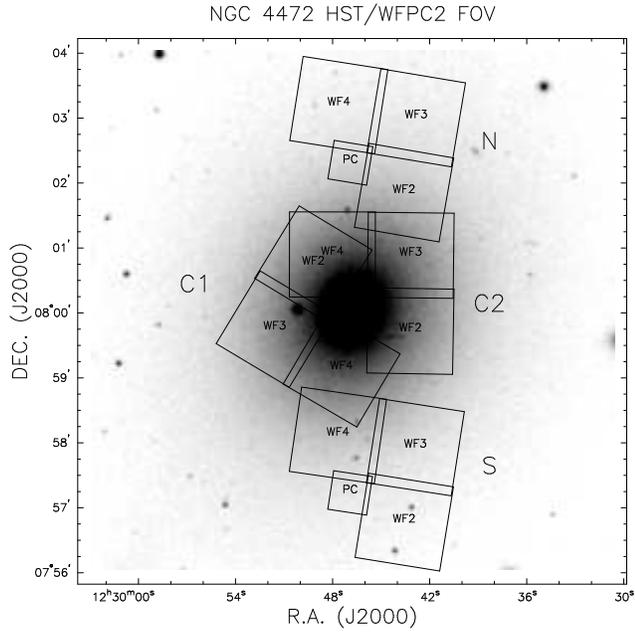}
\figcaption{Digitized Sky Survey image of the region showing positions and orientations
of the $HST$/WFPC2 fields of NGC 4472.
}
\end{figure}

\begin{figure}
\epsfig{figure=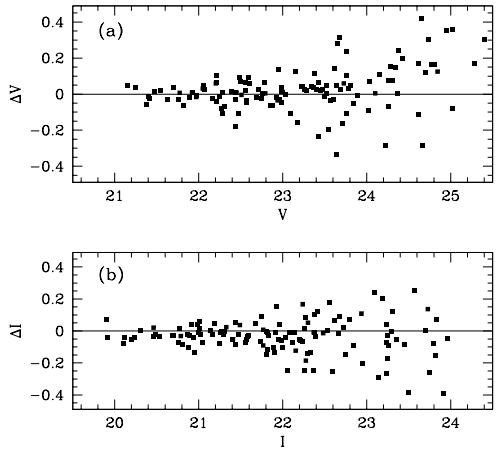, height=0.475\textwidth, width=0.475\textwidth}
\figcaption{Comparison of the photometry of the objects common in overlapped regions
between neighboring fields for $V$ (a) and $I$ (b).
}
\end{figure}

\begin{figure}
\epsfig{figure=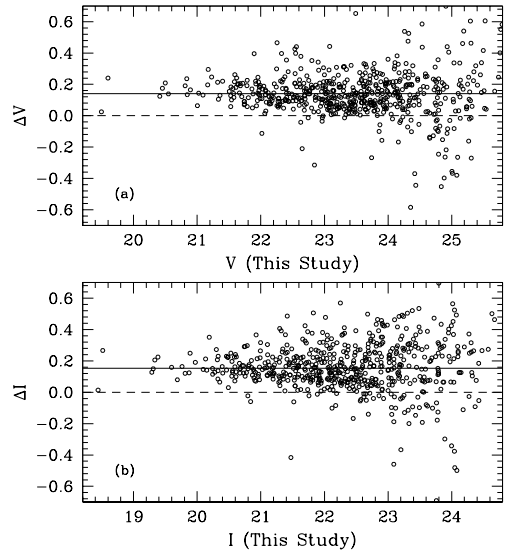, height=0.475\textwidth, width=0.475\textwidth}
\figcaption{ Comparison of the photometry of the objects common in this study
and Puzia et al.(1999). The differences are given in terms of this study minus
 Puzia et al.'s. The solid lines represent the median value of the differences.
}
\end{figure}

\begin{figure}
\epsfig{figure=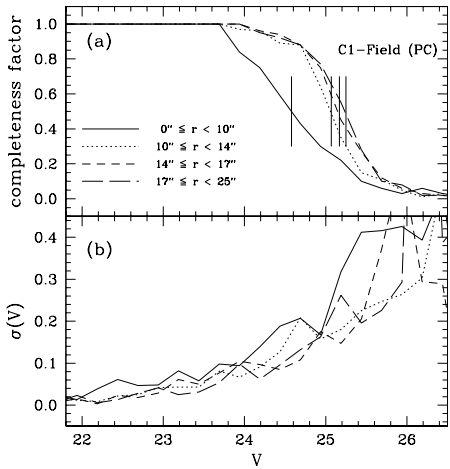, height=0.475\textwidth, width=0.475\textwidth}
\figcaption{
(a) Completeness of our photometry for the PC chip of the C1 Field, the
center of which corresponds almost the center of NGC 4472. Therefore the
incompleteness is expected to be the highest among the fields used in this
study. Different types of lines represent the regions at different radius
from the center of NGC 4472.
The vertical bars represent the level of 50\% completeness.
(b) Mean photometric errors versus magnitudes obtained from the artificial object experiment.
}
\end{figure}

\begin{figure}
\epsfig{figure=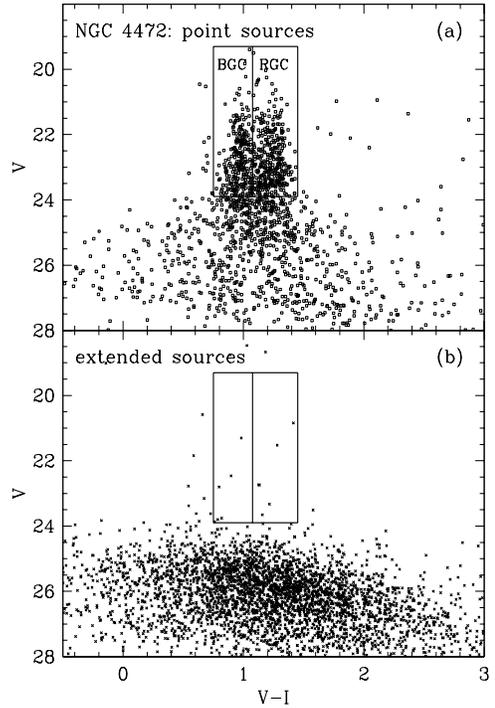, height=0.671\textwidth, width=0.475\textwidth}
\figcaption{
(a) $V$ vs. $(V-I)$ diagram of the measured point sources in the images of
NGC 4472.
The boundaries for the blue globular clusters (BGCs) and red globular clusters
(RGCs) brighter than $V=23.9$ mag are marked by the boxes.
(b) $V$ vs. $(V-I)$ diagram of the measured extended sources in the images of
NGC 4472.
 }
\end{figure}

\begin{figure}
\epsfig{figure=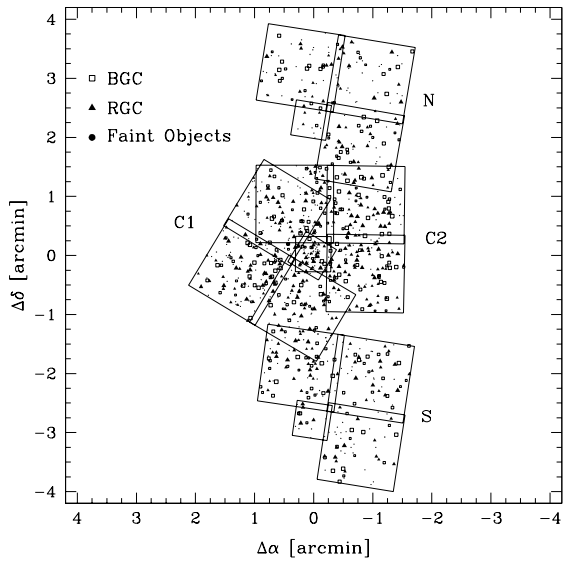, height=0.475\textwidth, width=0.475\textwidth}
\figcaption{
Positions of the globular cluster samples.
Open squares, filled triangles and dots represent, respectively,
the BGC with $V<23.9$ mag, RGC with $V<23.9$ mag and faint globular
clusters with $23.9<V<26$ mag.
}
\end{figure}

\begin{figure}
\epsfig{figure=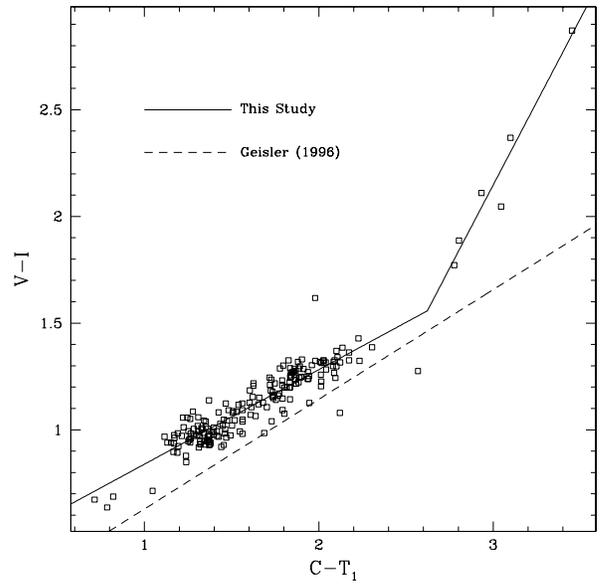, height=0.475\textwidth, width=0.475\textwidth}
\figcaption{
$(V-I)$ vs. $(C-T_1 )$ of the point sources in NGC 4472
common between this study and Geisler \etal (1996).
The solid line represents double-linear fits to the data and the dashed
line represents a transformation relation given by Geisler (1996).}
\end{figure}

\begin{figure}
\epsfig{figure=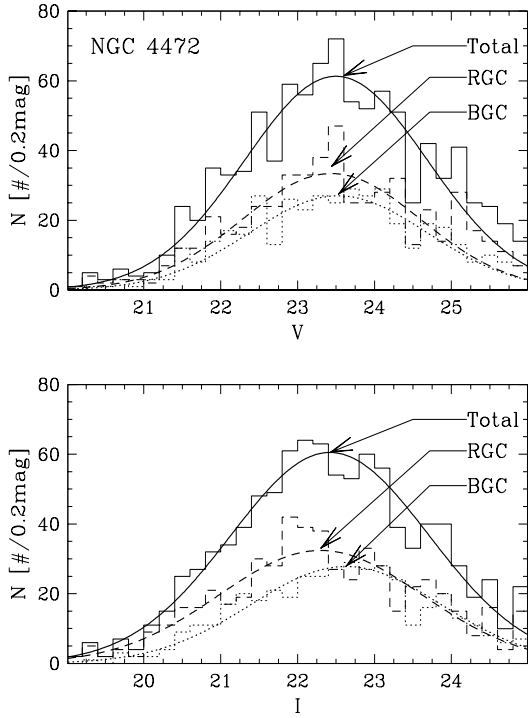, height=0.671\textwidth, width=0.475\textwidth}
\figcaption{
$V$-band and $I$-band luminosity functions of the
globular cluster candidates in NGC 4472 (histograms).
The solid line, dashed line, and dotted lines represent, respectively,
the luminosity functions for the total sample, RGCs, and BGCs.
The curved lines show Gaussian fits to the data.
}
\end{figure}

\begin{figure}
\epsfig{figure=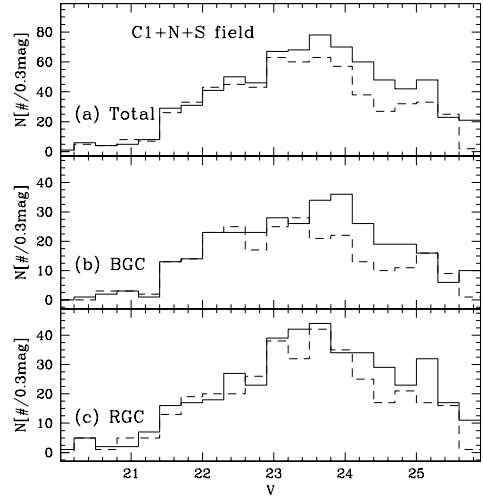, height=0.475\textwidth, width=0.475\textwidth}
\figcaption{
Comparison of the globular cluster luminosity functions (before
incompleteness correction) of this study
(the solid line) and Puzia et al. (1999) (the dashed line).
For the purpose of comparison, only C1, N and S
fields are used to derive the luminosity functions, as used by the latter.
Same color ranges are used for both samples: $0.75<(V-I)<1.08$ for the blue
 globular clusters and $1.08 <(V-I) < 1.45$ for the red globular clusters.
$V$ magnitudes of Puzia et al.'s are
adjusted by +0.14 mag to match our photometry.
Note that there are somewhat more objects at $V>23.5$ mag in this study than in
Puzia et al., showing that our sample is more complete than the latter.}
\end{figure}

\begin{figure}
\epsfig{figure=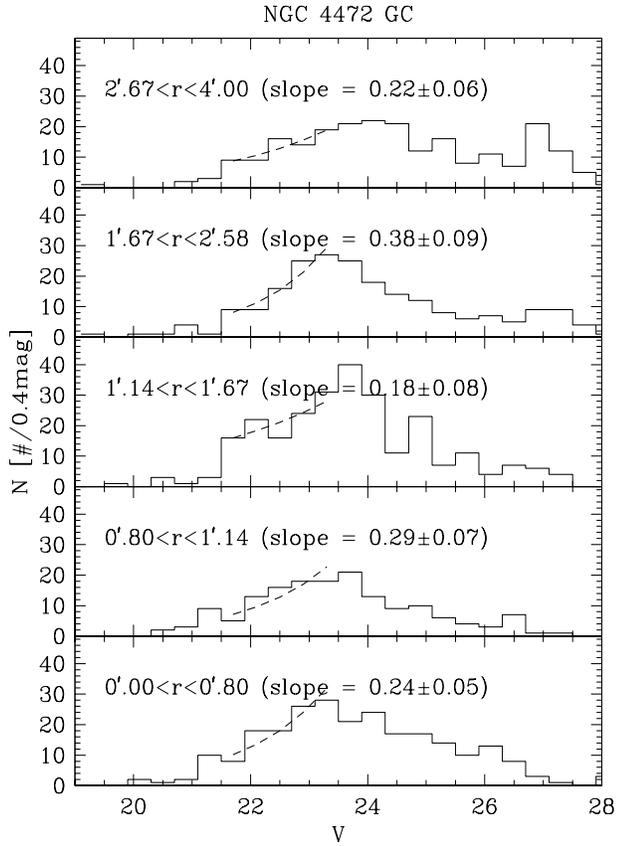, height=0.671\textwidth, width=0.475\textwidth}
\figcaption{
Radial variation of the globular cluster luminosity function for NGC 4472.
The dashed lines represent logarithmic fits to the data for the
range of $21.5<V<23.5$ mag. Note that there is little systematic change
in the slope of the fitted data as the galactocentric distance changes.
}
\end{figure}

\begin{figure}
\epsfig{figure=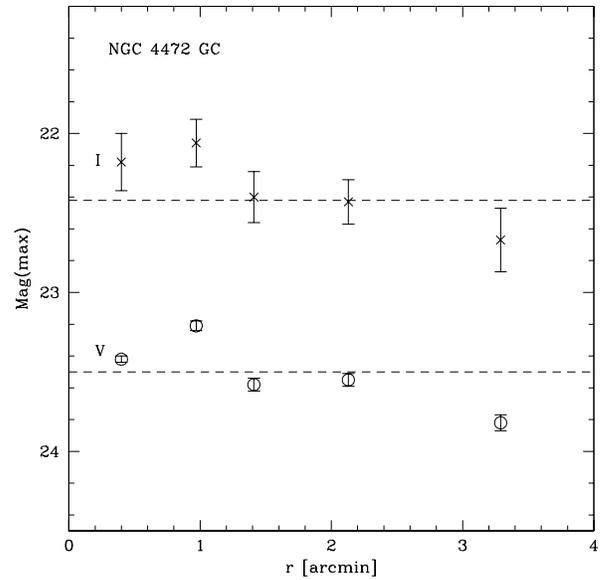, height=0.475\textwidth, width=0.475\textwidth}
\figcaption{
Radial variation of the peak luminosity of the globular cluster
luminosity functions for $V$ and $I$ bands.
The dashed lines represent the peak luminosity of the entire sample.
}
\end{figure}

\begin{figure}
\epsfig{figure=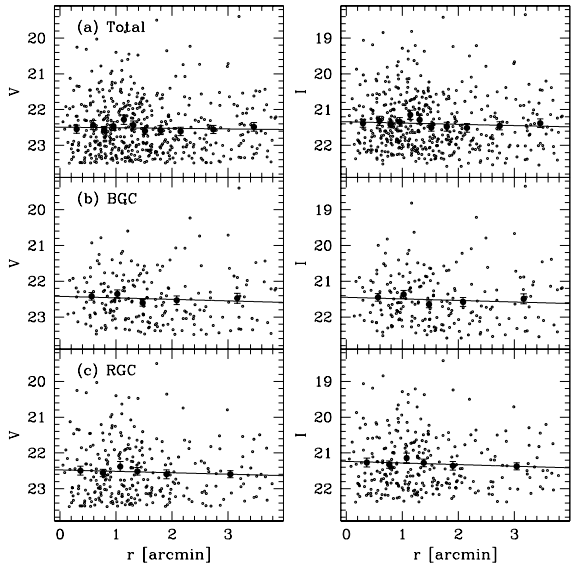, height=0.475\textwidth, width=0.475\textwidth}
\figcaption{
Radial variation of the mean magnitude of the bright globular clusters
with $V<23.5$ mag in the HST sample (the filled circles).
The solid lines represent linear fits to the mean magnitudes.
Note that there is little change in the mean magnitude with increasing
galactocentric distance.
}
\end{figure}

\begin{figure}
\epsfig{figure=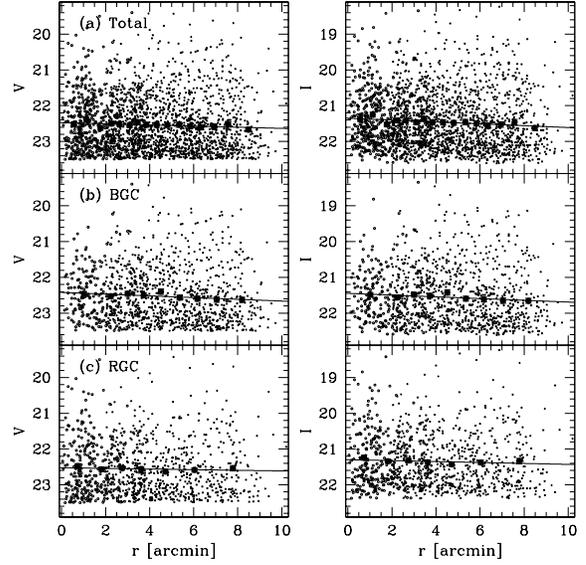, height=0.475\textwidth, width=0.475\textwidth}
\figcaption{
Radial variation of the mean magnitude of the bright globular clusters
with $V<23.5$ mag in the HST sample (the open circles) and
the ground-based sample given by Lee et al. (1998) (the crosses).
The solid lines represent linear fits to the mean magnitudes.
}
\end{figure}

\begin{figure}
\epsfig{figure=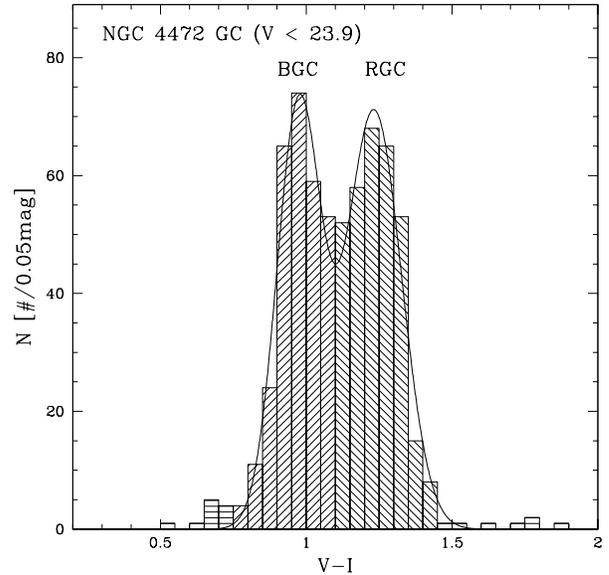, height=0.475\textwidth, width=0.475\textwidth}
\figcaption{
Color distribution of the bright globular clusters with $V<23.9$ mag
in NGC 4472. The curved lines show double-Gaussian fits to the data .}
\end{figure}

\begin{figure}
\epsfig{figure=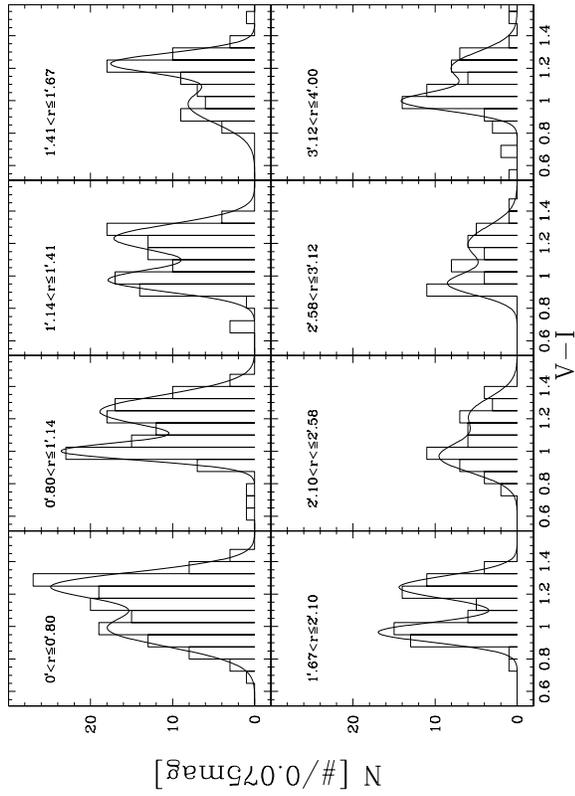, height=0.671\textwidth, width=0.475\textwidth}
\figcaption{
Radial variation of the color distribution of the globular clusters
with $V<23.9$ mag in NGC 4472.
The curved lines show double-Gaussian fits to the data.}
\end{figure}

\begin{figure}
\epsfig{figure=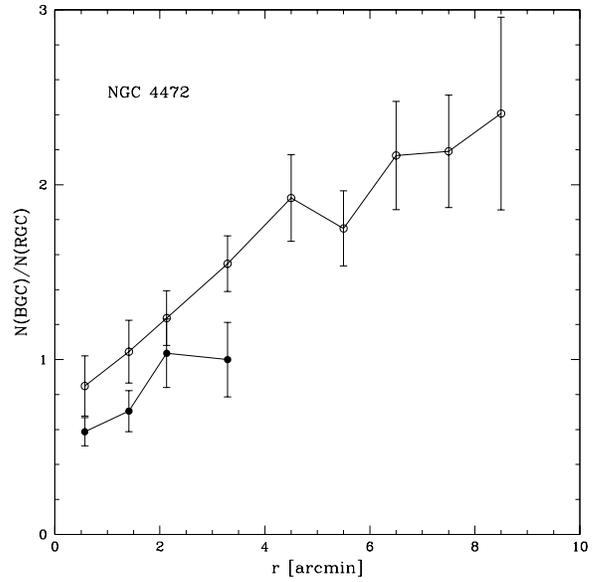, height=0.475\textwidth, width=0.475\textwidth}
\figcaption{
Radial variation of the number of the BGCs to that of the RGCs
with $V<23.9$ mag in NGC 4472 (filled circles).
The open circles represent the data for the globular clusters with
$T_1<23$ mag in the outer region of NGC 4472 derived from Geisler \etal (1996).}
\end{figure}

\begin{figure}
\epsfig{figure=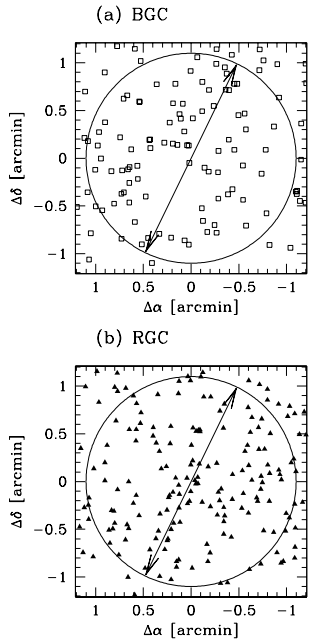, height=0.575\textwidth, width=0.575\textwidth}
\figcaption{
Spatial distribution of the blue (a) and red (b)
globular clusters brighter than $V=23.9$ mag in
the central region of NGC 4472. The radius of the circle is 1.1 arcmin
and the arrow represents the direction of the major axis of the stellar halo.}
\end{figure}

\begin{figure}
\epsfig{figure=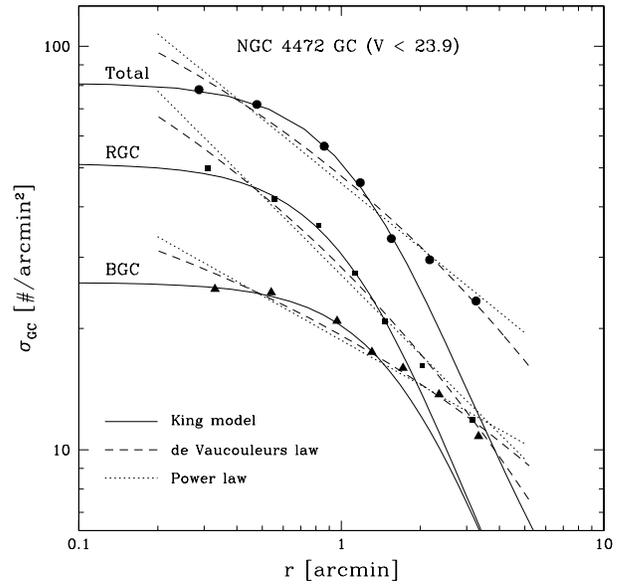, height=0.475\textwidth, width=0.475\textwidth}
\figcaption{
Radial variation of the surface number density of the bright globular clusters
with $V < 23.9$ mag
in the inner region of NGC 4472. The circles, triangles,  and squares
represent, respectively,
the total sample, blue globular clusters, and red globular clusters.
The solid line, dashed line and dotted line show, respectively,
fits to the data with King model, de Vaucouleurs law and power law.
}
\end{figure}
\clearpage

\begin{figure}
\epsfig{figure=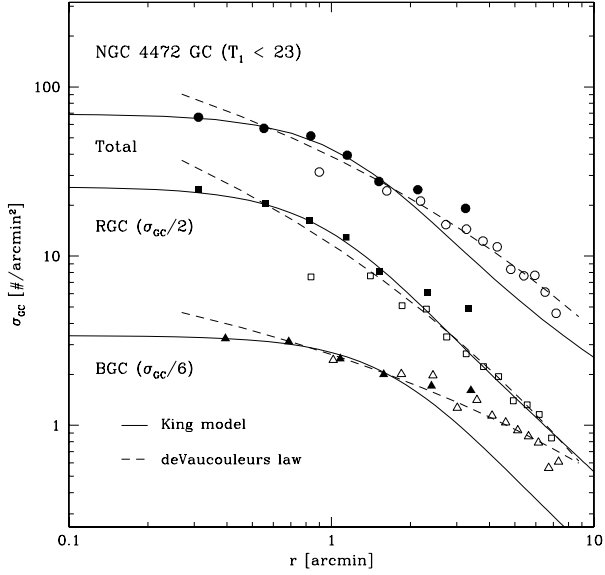, height=0.475\textwidth, width=0.475\textwidth}
\figcaption{
Radial variation of the surface number density of the bright globular clusters 
in the inner region ($V<23.7$ mag corresponding to $T_1 < 23.0$ mag, filled
symbols)
and the outer region ($T_1 < 23.0$ mag, open symbols)
of  NGC 4472 given by this study and Lee \etal (1998).
The circles, triangles, and squares represent, respectively,
the total sample, blue globular clusters, and red globular clusters.
The solid line and dashed line  show, respectively,
fits to the data with King model and de Vaucouleurs law.
The data for the RGCs and BGCs were scaled down, respectively,
 by a factor of 1/2 and 1/6 to show the data better.
}
\end{figure}

\begin{figure}
\epsfig{figure=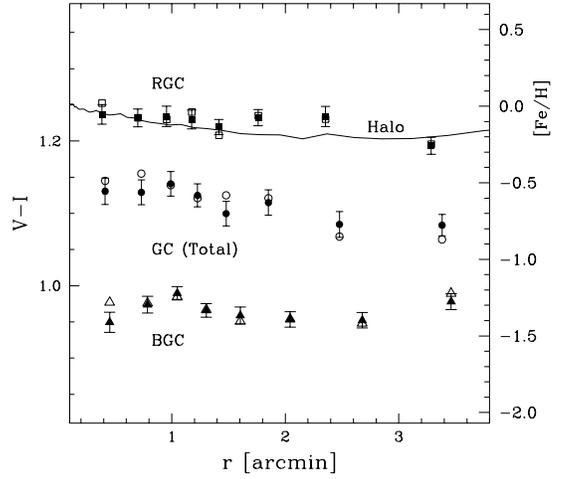, height=0.475\textwidth, width=0.475\textwidth}
\figcaption{
Radial variation of the mean (filled symbols) and median (open symbols)
color of the bright globular clusters
with $V<23.9$ mag in NGC 4472.
The  circles, triangles, and squares represent, respectively,
the total sample, blue globular clusters, and red globular clusters.
The solid line represents the mean color of the stellar halo of NGC 4472
given by Kim, Lee \& Geisler (2000).
}
\end{figure}

\begin{figure}
\epsfig{figure=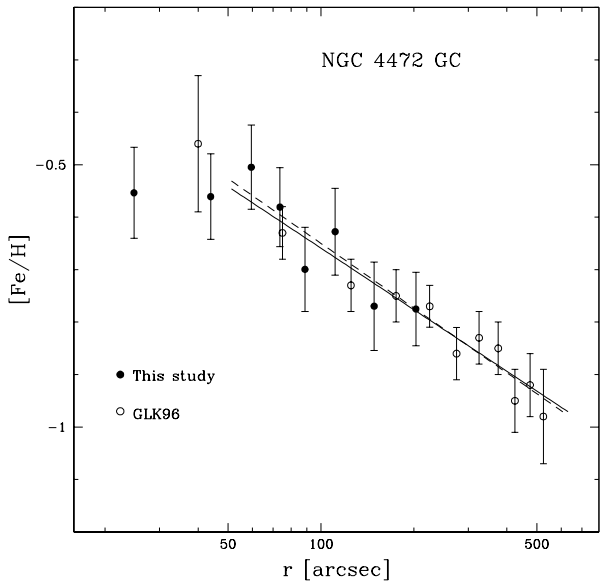, height=0.475\textwidth, width=0.475\textwidth}
\figcaption{Radial variation of the mean metallicity of the bright
globular clusters in the wide region of NGC 4472.
The filled and open circles represent, respectively, the data in this
study and those given by Geisler et al. (1996).
The solid line   shows a linear fit to the combined data of two studies,
and the dashed line represents a linear fit given by Geisler et al. (1996).
}
\end{figure}

\begin{figure}
\epsfig{figure=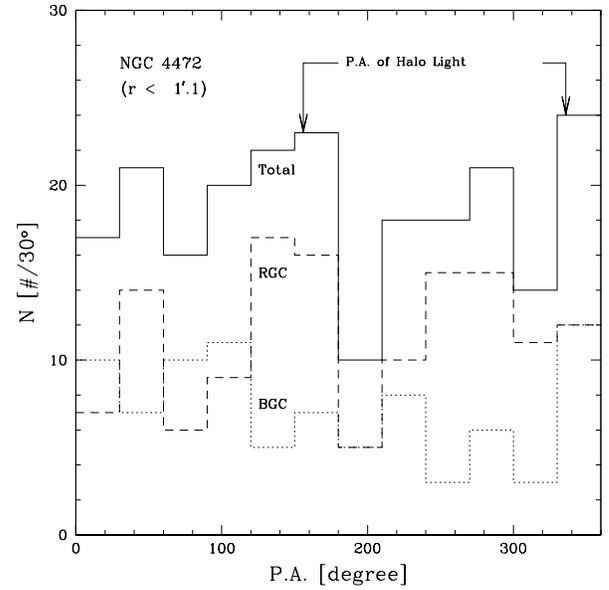, height=0.475\textwidth, width=0.475\textwidth}
\figcaption{
Azimuthal variation of the surface number density of the bright
globular clusters with $V<23.9$ mag in the region at $r<1.1$ arcmin.
The solid line, dotted line and dashed line show, respectively, the total
sample, blue globular clusters, and red globular clusters.
}
\end{figure}

\clearpage

\begin{deluxetable}{cccccc}
\footnotesize
\tablecaption{JOURNAL OF OBSERVATIONS FOR NGC 4472 IN THE HST ARCHIVE \label{tbl-1}}
\tablehead{
\colhead{Field} & \colhead{T$_{\rm exp}$(F555W)}   & 
\colhead{T$_{\rm exp}$(F814W)}   & Date & Program ID and PI}
\startdata
C1 Field & $2\times 900$ s & $2\times 900$ s & 1995 Feb. 4,5  & GO.5236 Westphal \nl
C2 Field & $2\times 230$ s & $2\times 230$ s & 1995 Apr. 18 & GO.5236 Westphal \nl
N Field &  900 s, 1300 s & 900 s, 1400 s & 1996 May 3 & GO.5920 Brodie \nl
S Field &  900 s, 1300 s & 900 s, 1400 s & 1996 May 2 & GO.5920 Brodie \nl
\enddata
 
\end{deluxetable}


\begin{deluxetable}{ccc}
\footnotesize
\tablecaption{APERTURE CORRECTIONS \label{tbl-2}}
\tablehead{
\colhead{Field, Image} & \colhead{APC(F555W)}   & 
\colhead{APC(F814W)} }
\startdata
C1 Field & & \nl
PC & $-0.434\pm0.047$ & $-0.626\pm0.099$ \nl
WF2 & $-0.202\pm0.020$ & $-0.257\pm0.033$ \nl
WF3 & $-0.213\pm0.014$ & $-0.261\pm0.019$ \nl
WF4 & $-0.182\pm0.017$ & $-0.239\pm0.026$ \nl
C2 Field & & \nl
PC & $-0.522\pm0.128$ & $-0.658\pm0.259$ \nl
WF2 & $-0.190\pm0.024$ & $-0.252\pm0.051$ \nl
WF3 & $-0.217\pm0.028$ & $-0.289\pm0.030$ \nl
WF4 & $-0.195\pm0.048$ & $-0.225\pm0.046$ \nl
N Field & & \nl
PC & $-0.506\pm0.022$ & $-0.541\pm0.015$ \nl
WF2 & $-0.205\pm0.022$ & $-0.244\pm0.015$ \nl
WF3 & $-0.206\pm0.016$ & $-0.280\pm0.078$ \nl
WF4 & $-0.267\pm0.072$ & $-0.314\pm0.093$ \nl
S Field & & \nl
PC & $-0.667\pm0.056$ & $-0.851\pm0.035$ \nl
WF2 & $-0.259\pm0.051$ & $-0.263\pm0.034$ \nl
WF3 & $-0.251\pm0.023$ & $-0.292\pm0.023$ \nl
WF4 & $-0.245\pm0.021$ & $-0.263\pm0.022$ \nl
\enddata
\end{deluxetable}

\begin{deluxetable}{ccccc}
\footnotesize
\tablecaption{COMPARISON OF THE PHOTOMETRY FOR THE OVERLAPPED REGIONS \label{tbl-3}}
\tablehead{
\colhead{Field} & \colhead{$\Delta V$} & \colhead{N} & \colhead{$\Delta I$} & 
\colhead{N} \nl    }
\startdata
C1 Field(PC) minus C2 Field(PC) & $0.017\pm0.066$ & 10 & $0.024\pm0.032$ &10 \nl
C1 Field(WF2) minus C2 Field(WF4) & $-0.001\pm0.047$ & 28 & $-0.043\pm0.045$ &32 \nl
C1 Field(WF4) minus S Field(WF4) & $-0.078\pm0.061$ & 5& $0.009\pm0.057$ &5\nl
C2 Field(WF3) minus N Field(WF2) & $0.011\pm0.028$ & 10& $-0.028\pm0.027$ &11\nl
total objects & $0.001\pm0.049$ & 52& $-0.026\pm0.051$ & 59 \nl
\enddata
 
\end{deluxetable}

\begin{deluxetable}{cccccc | cccccc }
\scriptsize
\tablecaption{PHOTOMETRY OF BRIGHT GLOBULAR CLUSTERS WITH $V<22$ MAG
IN NGC 4472 \label{tbl-4}}
\tablewidth{0pt}
\tablehead{
\colhead{ID} & \colhead{$\Delta RA('')^a$} & \colhead{$\Delta Dec('')^a$} & \colhead{$V$} & 
\colhead{$(V-I)$} &         \colhead{$r('')$} &   
\colhead{ID} & \colhead{$\Delta RA('')^a$} & \colhead{$\Delta Dec('')^a$} & \colhead{$V$} & 
\colhead{$(V-I)$}           & \colhead{$r('')$ }  
}
\startdata
  1061&  -91.3&  -30.6&  21.89&   1.32&  96.2&    90&    8.3& -109.6&  21.58&   0.89& 109.9\nl
   617&  -86.4&   14.6&  21.70&   1.25&  87.7&  1246&   12.0&  -44.8&  21.65&   1.20&  46.4\nl
   911&  -78.6&  -12.8&  21.26&   1.21&  79.6&    39&   14.6& -138.1&  21.69&   1.33& 138.9\nl
   741&  -73.8&   -1.8&  20.45&   1.24&  73.8&    23&   19.4& -122.4&  21.07&   0.85& 124.0\nl
   236&  -69.3&   21.0&  21.87&   1.16&  72.4&   179&   19.6& -218.8&  21.43&   1.05& 219.7\nl
  1062&  -68.0&  -44.6&  21.83&   1.18&  81.3&   371&   20.6&  -25.1&  21.45&   1.26&  32.5\nl
   398&  -67.0&   12.8&  21.82&   1.07&  68.2&   514&   25.6&   65.0&  19.82&   1.01&  69.9\nl
   417&  -64.8&   10.8&  21.22&   0.99&  65.6&   238&   25.7&  130.9&  21.64&   1.15& 133.4\nl
   833&  -64.5&  -14.7&  20.90&   1.31&  66.2&   584&   26.0&  -19.5&  21.62&   0.93&  32.5\nl
  1129&  -64.1&  -63.7&  21.57&   0.93&  90.4&   295&   28.9&  110.6&  20.35&   1.12& 114.3\nl
   705&  -60.2&   -7.2&  21.85&   0.99&  60.6&   542&   29.2&  211.9&  21.49&   1.29& 213.9\nl
   972&  -55.9&  -32.7&  21.49&   0.97&  64.8&   263&   31.1& -193.2&  21.54&   1.25& 195.7\nl
   233&  -54.1&   12.0&  21.30&   1.27&  55.4&   723&   32.7&  -26.2&  21.25&   1.04&  41.9\nl
   925&  -51.7&  -30.5&  20.40&   1.12&  60.0&   243&   33.7&   28.1&  20.78&   0.96&  43.8\nl
   772&  -50.0&  -18.1&  21.47&   1.11&  53.1&   541&   34.3&   75.8&  21.54&   1.00&  83.2\nl
   749&  -47.6&  202.2&  21.62&   1.17& 207.7&   512&   36.2&   63.6&  21.77&   1.08&  73.2\nl
  1153&  -46.3&   68.3&  21.64&   1.15&  82.5&   388&   38.3&   40.4&  20.93&   1.27&  55.6\nl
  1084&  -42.4&   37.5&  21.77&   0.98&  56.6&   305&   38.7& -177.3&  21.64&   0.92& 181.5\nl
   256&  -41.2&  -36.6&  21.69&   1.30&  55.1&   417&   40.7&   95.1&  21.60&   1.09& 103.5\nl
   314&  -39.3&   -0.1&  20.03&   1.19&  39.3&   327&   42.8& -124.8&  20.74&   1.25& 131.9\nl
   808&  -38.8&  -27.8&  21.76&   0.92&  47.8&   824&   43.8&   -4.8&  21.21&   1.12&  44.1\nl
   648&  -38.8&  -15.6&  21.51&   0.95&  41.8&   379&   44.4&   39.4&  21.26&   1.32&  59.3\nl
   480&  -38.0& -128.3&  21.07&   0.98& 133.8&   503&   45.5&   61.5&  21.43&   0.95&  76.5\nl
  1104&  -36.2&   54.0&  21.43&   1.28&  65.0&   549&   48.4&   79.5&  21.76&   0.97&  93.1\nl
   617&  -35.0&  188.5&  19.39&   1.05& 191.7&   407&   50.3& -179.5&  21.85&   0.96& 186.4\nl
   450&  -34.4&   -8.7&  21.78&   1.20&  35.4&   432&   51.0&  140.6&  20.74&   0.95& 149.6\nl
   347&  -33.8&   -4.7&  20.92&   0.93&  34.1&    99&   52.1&   20.0&  21.40&   0.99&  55.8\nl
   402&  -26.6& -106.2&  21.68&   1.25& 109.5&   368&   55.2&   38.1&  21.16&   1.03&  67.1\nl
   467&  -19.5&  -20.3&  20.31&   1.13&  28.2&   259&   61.4&  159.9&  21.88&   1.21& 171.3\nl
   322&  -18.7&    8.1&  21.35&   1.15&  20.4&   872&   61.8&  -22.7&  21.52&   1.14&  65.8\nl
  1198&  -18.4&  -86.6&  21.40&   1.18&  88.5&   556&   64.5&  136.9&  21.56&   0.95& 151.4\nl
   846&  -16.5&  -42.6&  21.44&   1.22&  45.7&   881&   65.9&    6.4&  20.63&   1.27&  66.2\nl
   317&  -16.2& -103.2&  21.54&   0.97& 104.4&   270&   68.0&   29.7&  21.58&   1.16&  74.2\nl
   179&  -15.6&    9.5&  20.25&   1.19&  18.2&   142&   69.2&   21.8&  20.59&   0.97&  72.6\nl
   441&  -13.1& -170.3&  21.52&   1.14& 170.8&   554&   69.9&   81.4&  21.70&   0.99& 107.3\nl
   766&  -12.5&  -30.1&  21.87&   1.08&  32.6&   427&   71.0&   43.9&  20.48&   1.11&  83.5\nl
   905&   -8.0&   55.3&  21.23&   1.09&  55.9&   312&   74.2&  164.1&  20.71&   1.05& 180.1\nl
   683&   -7.6&  -16.5&  21.34&   1.31&  18.1&   901&   74.4&  -43.8&  21.78&   0.91&  86.3\nl
    60&   -7.5&   -7.9&  21.13&   1.26&  10.9&   902&   75.9&    0.3&  21.08&   1.36&  75.9\nl
   344&   -6.8&  190.0&  21.89&   1.01& 190.1&   568&   76.1&   89.9&  21.40&   0.94& 117.8\nl
  1289&   -5.6&  -87.7&  21.85&   1.37&  87.9&   419&   76.5&   43.0&  21.68&   1.28&  87.7\nl
  1257&   -3.9&  -74.8&  21.88&   1.25&  74.9&   555&   79.5&  206.6&  21.86&   1.12& 221.3\nl
   413&    0.2&   22.9&  21.86&   1.26&  22.9&   389&   79.8&   39.9&  21.68&   1.04&  89.2\nl
  1199&    0.7&  -54.3&  20.99&   1.22&  54.3&   507&   82.6& -110.7&  21.88&   1.22& 138.1\nl
   339&    3.1&   31.6&  21.46&   1.30&  31.8&   911&   83.9&  -11.7&  21.83&   0.98&  84.7\nl
  1366&    4.3& -103.6&  19.50&   1.08& 103.6&   284&   85.4& -137.2&  21.82&   1.32& 161.6\nl
   292&    4.6&   32.3&  21.82&   1.27&  32.6&   499&   85.5&  191.5&  21.48&   1.02& 209.8\nl
    79&    6.9&  -12.9&  21.46&   1.16&  14.6&   523&   86.5& -109.2&  20.23&   1.02& 139.3\nl
  1182&    7.4&  -39.4&  21.43&   0.92&  40.1&   265&   88.7&  155.9&  20.79&   1.08& 179.3\nl
\enddata
\tablenotetext{a}{$\Delta RA$, $\Delta Dec$, and $r$ are calculated in the unit of arcsec
with respect to the center of NGC 4472, for the position angle of the major axis of NGC 4472, 58.67 deg. }
 \end{deluxetable}

\begin{deluxetable}{ccccc|cccc}
\footnotesize
\tablecaption{LUMINOSITY FUNCTIONS OF THE GLOBULAR CLUSTERS IN NGC 4472\label{tbl-5}}
\tablewidth{0pt}
\tablehead{
\colhead{$V$} & \colhead{N(total)} & \colhead{N(BGC)} &\colhead{N(RGC)} & &\colhead{$V$} & \colhead{N(total)} & \colhead{N(BGC)} &\colhead{N(RGC)}} 
\startdata
19.3 & 1 & 1 & 0 && 23.7 & 54 & 28 & 26 \nl
19.5 & 1 & 0 & 1 && 23.9 & 52 & 27 & 25 \nl
19.7 & 0 & 0 & 0 && 24.1 & 57 & 28 & 29 \nl
19.9 & 1 & 1 & 0 && 24.3 & 51 & 19 & 32 \nl
20.1 & 1 & 0 & 1 && 24.5 & 25 & 12 & 13 \nl
20.3 & 5 & 1 & 4 && 24.7 & 42 & 19 & 23 \nl
20.5 & 3 & 1 & 2 && 24.9 & 32 & 18 & 14 \nl
20.7 & 6 & 3 & 3 && 25.1 & 41 & 13 & 28 \nl
20.9 & 4 & 1 & 3 && 25.3 & 25 &  8&  17\nl
21.1 & 5 & 3 & 2 && 25.5 & 24 &  10& 14 \nl
21.3 & 10 &3  &  7&& 25.7 &19 &  8&  11\nl
21.5 & 24 &12  & 12 && 25.9 & 14 & 7 & 7 \nl
21.7 & 20 & 8 & 12 && 26.1 & 20 & 10 & 10 \nl
21.9 & 35 & 13 & 22 && 26.3 & 18 & 8 &  10\nl
22.1 & 33 &  17& 16 && 26.5 & 19 & 12 &  7\nl
22.3 &34  & 16 & 18 && 26.7 & 16 &  7&  9\nl
22.5 & 51 &  27& 24 && 26.9 & 19 &  8& 11 \nl
22.7 & 37 &  13&  24&& 27.1 & 23 &  9& 14 \nl
22.9 & 59 &  24& 35 && 27.3 & 10 &  4& 6 \nl
23.1 & 56 &  23& 33 && 27.5 &  8 &  6&2  \nl
23.3 & 65 &  27& 38 && 27.7 &  8 &  5& 3 \nl
23.5 & 72 &  25& 47 && 27.9 & 1  &  0& 1 \nl
\enddata 
\end{deluxetable}

\begin{deluxetable}{cccccccc}
\footnotesize
\tablecaption{LINEAR FITS OF MEAN MAGNITUDE PROFILES\label{tbl-6}}
\tablewidth{0pt}
\tablehead{
\multicolumn{2}{c}{SAMPLE} & \multicolumn{3}{c}{HST} & \multicolumn{3}{c}{HST+GROUND}\\
\multicolumn{2}{c}{} & \colhead{Slope} & \colhead{Intercept} & \colhead{$\sigma$} & \colhead{Slope} & \colhead{Intercept} & \colhead{$\sigma$} }
\startdata
Total  & $V$ & $0.016\pm0.039$ & $22.491\pm0.067$ & 0.091 & $0.016\pm0.007$ & $22.475\pm0.035$ & 0.052 \nl 
       & $I$ & $0.035\pm0.039$ & $21.341\pm0.067$ & 0.098 & $0.026\pm0.008$ & $21.350\pm0.035$ & 0.058 \nl \hline
BGC    & $V$ & $0.042\pm0.062$ & $22.421\pm0.105$ & 0.083 & $0.024\pm0.010$ & $22.419\pm0.050$ & 0.052 \nl
       & $I$ & $0.043\pm0.063$ & $21.449\pm0.106$ & 0.094 & $0.024\pm0.010$ & $21.447\pm0.051$ & 0.054 \nl \hline
RGC    & $V$ & $0.040\pm0.050$ & $22.474\pm0.089$ & 0.063 & $0.009\pm0.011$ & $22.523\pm0.050$ & 0.040 \nl
       & $I$ & $0.045\pm0.049$ & $21.236\pm0.088$ & 0.065 & $0.013\pm0.011$ & $21.297\pm0.050$ & 0.043 \nl
\enddata
\end{deluxetable}

\begin{deluxetable}{ccc|cc}
\footnotesize
\tablecaption{COLOR DISTRIBUTION OF THE GLOBULAR CLUSTERS WITH $V<23.9$ MAG
IN NGC 4472 \label{tbl-7}}
\tablewidth{0pt}
\tablehead{
\colhead{$(V-I)$} & \colhead{N} &  &  \colhead{$(V-I)$} & \colhead{N}    }
\startdata
0.625 & 1 & &1.125 & 68 \nl
0.675 & 5 & &1.175 & 58 \nl
0.725 & 4 & &1.225 & 68 \nl
0.775 & 4 & &1.275 & 65 \nl
0.825 & 11 && 1.325 & 53 \nl
0.875 & 24 & &1.375 & 15 \nl
0.925 & 65 & &1.425 & 8 \nl
0.975 & 74 & &1.475 & 1 \nl
1.025 & 59 & &1.525 & 1 \nl
1.075 & 53 & &1.575 & 0 \nl 
\enddata
 \end{deluxetable}

\begin{deluxetable}{cc | cc | cc}
\footnotesize
\tablecaption{SURFACE NUMBER DENSITY  OF THE GLOBULAR CLUSTERS WITH $V<23.9$ MAG
IN NGC 4472 \label{tbl-8}}
\tablewidth{0pt}
\tablehead{
\colhead{$r$[arcmin]} & \colhead{$\sigma_{GC}$(total)} & 
\colhead{$r$[arcmin]} & \colhead{$\sigma_{BGC}$} &  
\colhead{$r$[arcmin]} & \colhead{$\sigma_{RGC}$}     }

\startdata
0.29 & $78.11 \pm 12.35$ & 0.33 & $ 25.04\pm 6.07$ & 0.32 &$ 49.93\pm9.12 $ \nl
0.48 & $71.79 \pm 7.11$ & 0.54 & $24.56 \pm 3.67$ & 0.55 &$ 41.80 \pm 7.63$ \nl
0.86 & $56.56 \pm 5.60$ & 0.96 & $ 20.87\pm 3.11 $ & 0.82 & $ 25.96 \pm 4.72$ \nl
1.18 & $45.93 \pm 4.55$ & 1.31 & $ 17.46\pm 2.60$ & 1.13 & $ 27.41\pm 3.60$ \nl
1.55 & $33.38 \pm 3.31$ & 1.72 & $ 15.95\pm $ 2.38& 1.46 & $ 20.82\pm 2.73$ \nl
2.17 & $29.58 \pm 2.93$ & 2.36 & $ 13.72\pm 2.05$ & 2.04 & $ 16.21\pm 2.13$ \nl
3.26 & $23.37 \pm 2.35$ & 3.33 & $ 10.80\pm 1.75$ & 3.16 & $ 11.85\pm 1.58$ \nl
\enddata
\tablenotetext{a}{$\sigma_{GC}$ is given in units of \# arcmin$^{-2}$. }
\end{deluxetable}

\begin{deluxetable}{ccc | ccc | ccc}
\footnotesize
\tablecaption{COLOR PROFILES  OF THE GLOBULAR CLUSTERS WITH $V<23.9$ MAG
IN NGC 4472 \label{tbl-9} }
\tablecolumns{9}
\tablewidth{470pt}
\tablehead{
\multicolumn{3}{c}{Total} & \multicolumn{3}{c}{BGC} & \multicolumn{3}{c}{RGC} \\ 
\cline{1-3} \cline{4-6} \cline{7-9} \\
\colhead{$r$} & \colhead{$V-I$} & \colhead{$V-I$} &\colhead{$r$} & \colhead{$V-I$} &
 \colhead{$V-I$} & \colhead{$r$} & \colhead{$V-I$} & \colhead{$V-I$} \\
\colhead{(arcmin)} & \colhead{mean} & \colhead{median} &
\colhead{(arcmin)} & \colhead{mean} & \colhead{median} &
\colhead{(arcmin)} & \colhead{mean} & \colhead{median} }
\startdata
0.413 & $1.131\pm0.161$ & 1.145 & 0.454 & $0.949\pm0.080$ & 0.977 & 0.385 & $1.237\pm0.089$ & 1.253   \nl 
0.732 & $1.129\pm0.151$ & 1.155 & 0.786 & $0.974\pm0.066$ & 0.976 & 0.701 & $1.232\pm0.081$ & 1.232  \nl
0.991 & $1.140\pm0.149$ & 1.139 & 1.048 & $0.990\pm0.053$ & 0.985 & 0.955 & $1.234\pm0.093$ & 1.230 \nl
1.227 & $1.125\pm0.140$ & 1.121 & 1.304 & $0.966\pm0.054$ & 0.967 & 1.176 & $1.230\pm0.086$ & 1.240  \nl
1.479 & $1.100\pm0.149$ & 1.125 & 1.604 & $0.959\pm0.067$ & 0.951 & 1.415 & $1.219\pm0.070$ & 1.208  \nl
1.852 & $1.115\pm0.154$ & 1.121 & 2.043 & $0.953\pm0.061$ & 0.954 & 1.762 & $1.232\pm0.075$ & 1.235  \nl
2.477 & $1.085\pm0.157$ & 1.068 & 2.678 & $0.952\pm0.061$ & 0.948 & 2.355 & $1.234\pm0.092$ & 1.250 \nl
3.381 & $1.084\pm0.130$ & 1.064 & 3.462 & $0.978\pm0.064$ & 0.990 & 3.286 & $1.193\pm0.078$ & 1.196  \nl
\enddata
\end{deluxetable}


\begin{thebibliography}{}


\bibitem[Ashman \& Zepf 1992]{ash92} Ashman, K.M., \& Zepf, S. E. 1992, 
       ApJ, 384, 50 

\bibitem[Bertin \& Arnouts 1996]{ber96} Bertin, E., \& Arnouts, S. 1996,
A\&AS, 117, 393
\bibitem[Bruzual \& Charlot 1996]{bru96} Bruzual, A. G., \& Charlot, S. 1996,
electronically available. See Leitherer, C., et al. 1996, PASP, 108, 996 

\bibitem[Burstein \& Heiles 1982]{bur82} Burstein, D., \& Heiles, C. 1982, AJ, 87, 65

\bibitem[Forbes \etal 1997]{for97} Forbes, D. A., Brodie, J. P., \& Grillmair, C. J. 1997, AJ, 113, 1652 

\bibitem[Gebhardt \& Kissler-Patig 1999]{geb99} Gebhardt, K., \& Kissler-Patig, M. 1999, ApJ, 118, 1526

\bibitem[Geisler 1996]{gei96} Geisler, D. 1996, PASP, 111, 480

\bibitem[Geisler \etal 1996]{geietal96} Geisler, D., Lee, M. G., \& Kim, E. 1996,
AJ, 111, 1529

\bibitem[Gnedin 1997]{gne97} Gnedin, O. Y. 1997, ApJ, 487, 663

\bibitem[Grillmair \etal 1986] {gri86} Grillmair, C. J., Pritchet, C., \& van den Bergh,
S. 1986, AJ, 91, 1328

\bibitem[Harris 1996]{har96}Harris, W. E. 1996, AJ, 112, 1487

\bibitem[Harris \etal 1998]{har98} Harris, W. E., Harris, G. L. H., \& McLaughlin, D. E. 1998, ApJ, 115, 1801

\bibitem[Holtzman \etal 1995a]{hol95a} Holtzman, J. \etal (the WFPC2 team)
1995a, PASP, 107, 156

\bibitem[Holtzman \etal 1995b]{hol95b} Holtzman, J. \etal (the WFPC2 team)
1995b, PASP, 107, 1065

\bibitem[Kim \etal 1999]{kim99} Kim, E., Lee, M. G., \& Geisler, D. 2000, MNRAS,
 in press

\bibitem[King 1966]{kin66} King, I. 1966, AJ, 71, 276

\bibitem[Kron 1980]{kro80} Kron, R. 1980, ApJS, 43, 305
\bibitem[Kundu 1999]{kun99b} Kundu, A. 1999, Ph.D. Thesis, University of Maryland
\bibitem[Kundu \etal 1998]{kun98} Kundu, A., Whitmore, B. C. 1998, AJ, 116, 2841

\bibitem[Kundu \etal 1999]{kun99} Kundu, A., Whitmore, B. C., Sparks, W. B.,
Macchetto, F. D., Zepf, S. E., \& Ashman, K. M. 1999, ApJ, 513, 733

\bibitem[Kurth \etal 1999]{kur99} Kurth, O. M., Fritze-v. Alvensleben, U.,
\& Fricke, K. J. 1999, A\&AS, 138, 19

\bibitem[Lee \etal 1998]{lee98} Lee, M. G., Kim, E. \& Geisler, D. 1998, AJ, 115, 947
\bibitem[Maraston 1998]{mar98} Maraston, C. 1998, MNRAS, 300, 872

\bibitem[Mighell 1998]{mig98} Mighell, K. 1998, private communication

\bibitem[Murali \& Weinberg 1997a]{mur97a} Murali, C., \& Weinberg, M. D. 1997a,
MNRAS, 288, 749
\bibitem[Murali \& Weinberg 1997b]{mur97b} Murali, C., \& Weinberg, M. D. 1997b,
MNRAS, 288, 767
\bibitem[Murali \& Weinberg 1997c]{mur97c} Murali, C., \& Weinberg, M. D. 1997c,
MNRAS, 291, 717
\bibitem[Ostriker \& Gnedin 1997]{ost97} Ostriker, J. P., \& Gnedin, O. Y. 1997,
ApJ, 487, 667
\bibitem[Puzia \etal 1999]{puz99} Puzia, T. H., Kissler-Patig, M., Brodie, J. P., \& Huchra, J. P. 1999, AJ, 118, 2734

\bibitem[Schlegel, Finkbeiner, \& Davis 1998]{sch98} Schlegel, D. J., 
  Finkbeiner, D. P., \&     Davis, M. 1998, ApJ, 500, 525

\bibitem[Williams \etal 1996]{wil96} Williams, R.  et al. 1996, AJ, 112, 1335
\bibitem[Worhtey 1994]{wor94} Worthey, G. 1994, ApJS, 95, 107

\end{thebibliography}
\end{document}